\documentclass[fleqn,usenatbib]{mnras}

\usepackage{newtxtext,newtxmath}
\usepackage[T1]{fontenc}

\DeclareRobustCommand{\VAN}[3]{#2}
\let\VANthebibliography\thebibliography
\def\thebibliography{\DeclareRobustCommand{\VAN}[3]{##3}\VANthebibliography}

\usepackage{graphicx}	% Including figure files
\usepackage{amsmath}	% Advanced maths commands
\usepackage{subcaption}
\graphicspath{ {./figures/} }

\newcommand{\Fearth}{F_\oplus}
\newcommand{\bd}{\boldsymbol}

\title[M-R-F relation with Bayesian Mixture Model]{Predicting Exoplanet Mass from Radius and Incident Flux: A Bayesian Mixture Model}

\author[Q. Ma \& S. K. Ghosh]{
Qi Ma,$^{1}$\thanks{E-mail: qma4@ncsu.edu} and
Sujit K. Ghosh,$^{1}$
\\
$^{1}$Department of Statistics, North Carolina State University, 2311 Stinson Drive, Raleigh, NC 27695, USA\\
}

\date{Accepted XXX. Received YYY; in original form ZZZ}

\pubyear{2020}

\begin{document}
\label{firstpage}
\pagerange{\pageref{firstpage}--\pageref{lastpage}}
\maketitle

% Abstract of the paper
\begin{abstract}
The relationship between mass and radius (M-R relation) is the key for inferring the planetary compositions and thus valuable for the studies of formation and migration models. However, the M-R relation alone is not enough for planetary characterization due to the dependence of it on other confounding variables. This paper provides a non-trivial extension of the M-R relation by including the incident flux as an additional variable. By using Bayesian hierarchical modeling (BHM) that leverages the flexibility of finite mixture models, a probabilistic mass-radius-flux relationship (M-R-F relation) is obtained based on a sample of 319 exoplanets. We find that the flux has nonnegligible impact on the M-R relation, while such impact is strongest for hot-Jupiters. On the population level, the planets with higher level of flux tend to be denser, and high flux could trigger significant mass loss for plants with radii larger than $13R_{\oplus}$. As a result, failing to account for the flux in mass prediction would cause systematic over or under-estimation. With the recent advent of computing power, although a lot of complex statistical models can be fitted using Monte Carlo methods, it has largely remain illusive how to validate these complex models when the data are observed with large measurement errors. We present two novel methods to examine model assumptions, which can be used not only for the models we present in this paper but can also be adapted for other statistical models.
\end{abstract}


%%%%%%%%%%%%%%%%%%%%%%%%%%%%%%%%%%%%%%%%%%%%%%%%%%

%%%%%%%%%%%%%%%%% BODY OF PAPER %%%%%%%%%%%%%%%%%%

\section{Introduction} \label{sec:intro}
With hundreds of confirmed exoplanets whose measured masses and radii are well constrained by the \textit{Kepler} Mission and subsequent radial velocity campaigns, the M-R relation has been well studied on the population level in recent years \citep[e.g.][]{Wei14, Had14, Bas17, Wol16, Chen17, Nin18, kanodia2019mass, Ma19}. Such relation itself and the associated astrophysical scatter for different exoplanet populations are vital for inferring bulk compositions and understanding the planet formation pathways.

However, the M-R relation alone is not enough for planetary characterization since it could be affected by multiple other planetary properties such as the incident flux. It is widely accepted that the flux has nonnegligible impact on the M-R relation on both observational and theoretical grounds. For example, a large fraction of hot-Jupiters have inflated radii larger than expected by models of gas giants cooling and contraction \citep{Mil11, Dem11}. Such abnormality is found to be related to the stellar irradiation, and a variety of inflation models have been proposed \cite[e.g.][]{Arr10, Wu12, Tre17}. The mass loss driven by XUV photoevaporation could also contribute to the M-R relation for both hot-Jupiters and lower-mass planets \citep[e.g.][]{Lam03,Yel04, Lop12, Hub07}.

Ignoring the effect of flux could amplify the intrinsic scatter of the estimated M-R relation and thus limit our accurate interpretation of it. Therefore, the relation between the mass, radius and flux (M-R-F relation hereafter) has been explored recently. \citet{Wei13} studied a sample of 135 planets and identified a break point in the M-R-F relation at $150M_{\oplus}$ by visual inspection. They applied a simple power law to approximate the M-R-F relation and concluded that $(R/R_{\oplus}) = 1.78(M/M_{\oplus})^{0.53}(F/\text{erg s}^{-1}\text{cm}^{-2})^{-0.03}$ for planets with masses less than $150M_{\oplus}$, and $(R/R_{\oplus}) = 2.45(M/M_{\oplus})^{-0.039}(F/\text{erg s}^{-1}\text{cm}^{-2})^{-0.094}$ for heavier planets. Following the work by \citet{Wol16}, \citet{Nei18} explored the dependence of the M-R relation on flux and host star mass for small planets ($R < 8R_{\oplus}$) using a Bayesian hierarchical model. However, they found it difficult to distinguish the effects of flux and host star mass with their limited data set, and preferred the model only considering the host start mass by performing model selection via WAIC \citep{Wat13}. \citet{Ses18} used a sample of 286 gas giants to investigate how the mass and the flux influence the radius inflation. Their probabilistic model shows that the planets with masses between $0.37M_J$ and $0.98M_J$ exhibits the strongest correlation between the inflated radius and the flux.

In this work, we present a probabilistic M-R-F relation using Bayesian hierarchical modelling that  leverages the flexibility of finite mixture models. We also focus on model checking techniques that can be applied to other models easily.

\section{Motivating Data Set} \label{sec:data}

The data considered in this work were acquired from NASA Exoplanet Archive \citep{Ake13} on 09/08/2019. There are a total of 319 transiting exoplanets in our sample that satisfy the following criteria:

\begin{enumerate}
    \item They have radial velocity (RV) or transit timing variation (TTV) mass measurements. In addition, their measurements of orbital semi-major axis in astronomical units ($a$), stellar temperature ($T_{\star}$), and stellar radius ($R_{\star}$) are also required to be available, as the incident flux in Earth units is computed using

    \begin{equation}
    \frac{F}{\Fearth} = \left(\frac{T_{\star}}{T_\odot}\right)^4\left(\frac{R_{\star}}{R_\odot}\right)^2\left(\frac{1}{a}\right)^2
    \end{equation}

    where $T_\odot$ and $R_\odot$ are the effective temperature and radius of sun respectively.
    
    \item Following \citet{Wol16}, in case of asymmetric upper and lower error bars ($\sigma^{obs}_{+} \ne \sigma^{obs}_{-}$), the average $\sigma^{obs} = \frac{1}{2}(\sigma^{obs}_{+} + \sigma^{obs}_{-})$ is taken as the standard deviance of the measurement as discussed in section \ref{subsec:residual}. To obtain a sample with relatively high signal-to-noise ratio, a $3\sigma$ cutoff is then applied to all measurements, i.e., $M^{ob}/\sigma^{Mob} > 3$,$R^{ob}/\sigma^{Rob} > 3$, and  $F^{ob}/\sigma^{Fob} > 3$ \citep{Chen17}. 
    
    \item Since our work is focusing on exoplanets, we exclude brown dwarfs exhibiting deuterium fusion by introducing an upper mass boundary at $13M_J$ \citep{Spi11}.
\end{enumerate}

\begin{figure}
\centering
   \includegraphics[width=8cm]{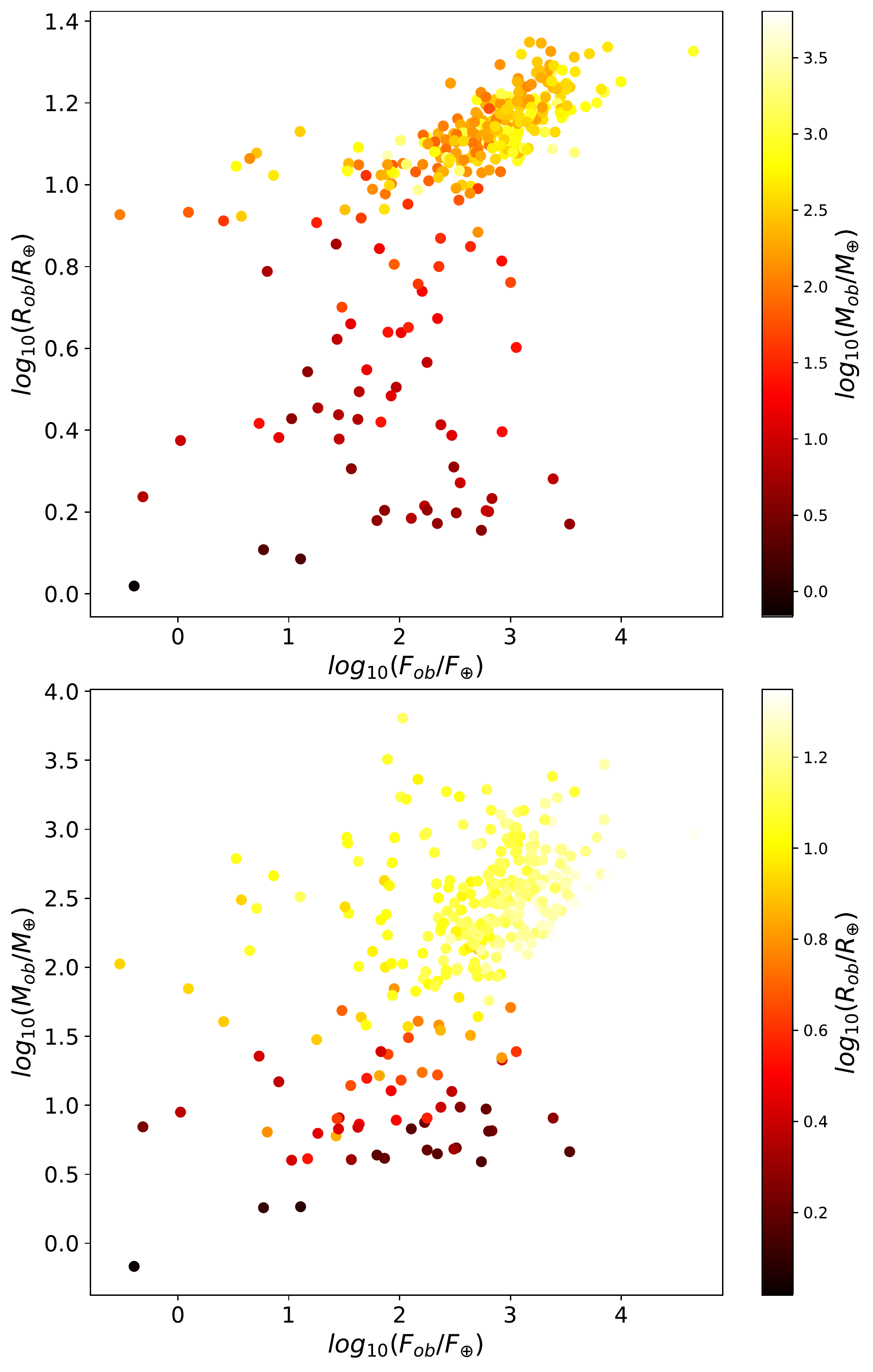}
   \caption{\small
   \textit{Top Panel}: Scatter plot of $\log F^{ob}$ vs $\log R^{ob}$ with $\log M^{ob}$ color-coded. \textit{Bottom Panel}:  Scatter plot of $\log F^{ob}$ vs $\log M^{ob}$ with $\log R^{ob}$ color-coded.}
   \label{fig:mrf_scatter}
\end{figure}

In this work, the mass, radius and flux measurements are all in earth units. Since they differ by several orders of magnitude, we also apply base-ten logarithmic transformation to them for numerical stability and efficiency of using MCMC sampling. For simplicity, we denote $\widetilde{M} = \log_{10}(M)$, $\widetilde{R} = \log_{10}(R)$, and $\widetilde{F} = \log_{10}(F)$. Figure \ref{fig:mrf_scatter} shows the distribution of the flux, radius and mass measurements in our sample. We latter use these plots to guide us developing joint models for mass, radius and flux. 

Unavoidably, the sample used in this work is perhaps subject to some level of selection bias that could impact the estimated M-R-F relation to some extent. Incorporating the selection bias into the model is certainly a critical step for more accurate estimation and can be performed using appropriately weighted version of likelihood \citep[e.g.][]{Ful17, Nei20}. However, as primary goals of this work, we mainly focus on (i) how to fit the joint distribution on heterogeneous data using a mixture model; and (ii) how to verify various modeling assumptions made via the specification of prior and sampling distributions. One of the challenging aspects of the developing model verification methods is how to deal with data that are measured with errors. We also note that RV and TTV measurements exhibit different observational bias with TTV characterized planets typically having lower densities \citep{Mil17, Ste16}. Following the previous works \citep[e.g.][]{Wol16, Nin18}, the inclusion of both techniques provides more data points especially in the sparse region of the M-R-F parameter space and thus could makes the inferred M-R-F relation more representative.

\section{Statistical Models} \label{sec:model}
A quick look at top panel in Figure \ref{fig:mrf_scatter} indicates that the joint probability distribution of radius and flux may depend on the magnitude of the masses. The bottom panel indicates that the joint distribution of mass and flux may depend on the magnitudes of radii (via two or three clusters). Thus, mixture models that allow us to capture the variations (clusters) of the joint distributions of two objects when varies with the level of a third object seem appropriate. In this section, we first provide a brief overview of some basic concepts of finite mixture models (FMM) and then provide details of our Bayesian hierarchical mixture model (BHMM) guided by the observed scatter plots in Figure \ref{fig:mrf_scatter}.

\subsection{FMM in Astronomy}

The probability density of a real-valued random variable modeled using an $m$-component FMM takes the form
\begin{equation}
    f(x; \bd{\Phi}) = \sum_{i=1}^m \pi_i g_i(x; \bd{\theta}_i),
    \label{eq:fmm}
\end{equation}
where $g_i(x; \bd{\theta}_i)$ is the $i$th component density known up to a parameter vector $\bd{\theta}_i$, $\{\pi_i\}_{i=1}^m$ are the nonnegative mixture weights that sum to $1$, and $\bd{\Phi} = (\pi_1, \ldots, \pi_m, \bd{\theta}_1, \ldots, \bd{\theta}_m)$ denotes the vector consisting of all the unknown parameters of the FMM.

For convenience with computational inference and better understanding of the mixture model, we can also express FMM in a hierarchical form. Let $Z$ be a categorical random variable taking values $1$ to $m$ with probabilities $\pi_1, \ldots, \pi_m$, respectively. Then, the distribution of the random variable $X$ with density given in equation (\ref{eq:fmm}), denoted by $X \sim f(x; \bd{\Phi})$, is equivalent to
\begin{equation}
    X|Z = i \sim g_i(x; \bd{\theta}_i), \; i = 1, \ldots, m.
\label{eq:mixgen}
\end{equation}
Equation (\ref{eq:mixgen}) describes the generating process of the random variable $X$ conditioned on the latent component indicator variable $Z$. The component densities $g_i$ are often chosen as Gaussian densities with mean $\mu_i$ and standard deviation $\sigma_i$, in which case $\bd{\theta}_i=(\mu_i, \sigma_i)$, but other probability densities (e.g., location-scale families like Cauchy, Laplace etc.) can also be useful in some settings depending on the range of the random variables \citep[e.g.][]{Pee00, Lee16}.

FMM are a natural choice for many applications in astronomy mainly for two reasons. First of all, FMM provides conceptually and computationally convenient way to model the data exhibiting group-structure, and is thus exploited for classification and cluster analysis of astronomical objects. For example, \citet{Lee12} applies a 6-component Gaussian mixture model to the pulsar distribution in the period-period derivative space, and identifies two possible clusters of millisecond pulsars. \citet{Ein12} studies multimodality in galaxy clusters using multivariate Gaussian mixture modelling, and shows the presence of complex substructures.

In addition, under some very mild regularity conditions, FMM as a flexible and semi-parametric approach can be used to approximate any unknown continuous density by choosing appropriately large $m$ in terms of total variation norm \citep{Roe97, Li00, nguyen2020approximation}. FMM is also better equipped with avoiding the common problem of choosing restrictive parametric probability densities (e.g. normal, log-normal, or gamma distributions) which can lead to considerable discrepancy between the chosen parametric model and the data, particularly when such astronomical data arise from (often unknown) complex astrophysical processes. Besides serving as an approximation of the target probability distribution directly \citep[e.g.][]{Mel18}, FMM can be used effectively for deriving or approximating a complex model using a sequence of simpler hierarchical models guided by the insights learned from 2-d scatter plots \citep[e.g.][]{Kel07}.

\subsection{The Problem of Label Switching}
In making statistical inference, finite mixture models may suffer from the problem of label switching due to the invariance of likelihood to the permutation of model parameters, and it is well-known that even maximum likelihood estimates don't exist without some restriction (e.g., see Section 3.10 of \citep{Pee00} for detail discussions). Suppose that there is a mixture model consisting of only two normal components with different means:

\begin{equation}
    f(x|w, \mu_1, \mu_2) = w\mathcal{N}(x|\mu_1,\sigma) + (1-w)\mathcal{N}(x|\mu_2,\sigma)
\label{eq:label_swicth}
\end{equation}

It's easy to verify that $f(x|w,\mu_1,\mu_2) = f(x|1-w, \mu_2, \mu_1)$. In other words, the likelihood would be \textit{invariant} and two distinct values of the parameters and thus it's hard to identify the model parameters through the likelihood of the data. The problem gets even worse as the number of mixture components grows, because there would be more permutations of model parameters that lead to the same likelihood value. Hoever, if we are interested in estimating the density (and its parameters), this problem is not much of a concern.

Label switching doesn't affect posterior predictive inference as all of the integrals involved are not affected by the ordering of indices of the components. However, inferences about the parameters are problematic under label switching, including the posterior estimates of individual parameters. For example, the posterior mean of $w$ in Equation \ref{eq:label_swicth} may always be close to $0.5$ with any data. Moreover, label switching usually results in highly multimodal posteriors that makes all known MCMC sampling techniques inefficient due to poor mixing of the chains (see Chapter 22.3 of \citet{Gel13} for more details).

The most common approach to deal with label switching is to impose ordering constraints on the parameters that identifies the component. For example, for the model describe in Equation \ref{eq:label_swicth}, a possible constraint is $\mu_1 < \mu_2$ that forbids the swapping of $\mu_1$ and $\mu_2$ making the likelihood informative about these parameters. Other alternatives  are to impose $w\geq 0.5$ and if we allow different $\sigma_1$ and $\sigma_2$ for the components, we can impose $\sigma_1<\sigma_2$ as well. All of these order restriction techniques will be applied in our modeling as described in Section \ref{subsubsec:marginal_f} and \ref{subsubsec:cond_R}.

\subsection{BHMM for the M-R-F Relation}

Our BHMM is designed to infer the joint density of the \textit{true unobserved} mass, radius and flux of a planet, denoted as $f(M, R, F)$, which can be factorized as

\begin{equation}
    f(\widetilde{M}, \widetilde{R}, \widetilde{F}) = f(\widetilde{F})f(\widetilde{R}|\widetilde{F})f(\widetilde{M}| \widetilde{R}, \widetilde{F}).
    \label{eq:joint}
\end{equation}
Other possible equivalent factorizations of the joint density in terms of appropriate conditional and marginal densities are also possible, but we use the above form for the convenience of our model development. In this section, we describe the details of modelling the three conditional density components on the right side of Equation (\ref{eq:joint}), as well as how the measurement errors are incorporated into the model.

\subsubsection{Marginal Distribution of Flux}\label{subsubsec:marginal_f}
\begin{figure}
\centering
   \includegraphics[width=7cm]{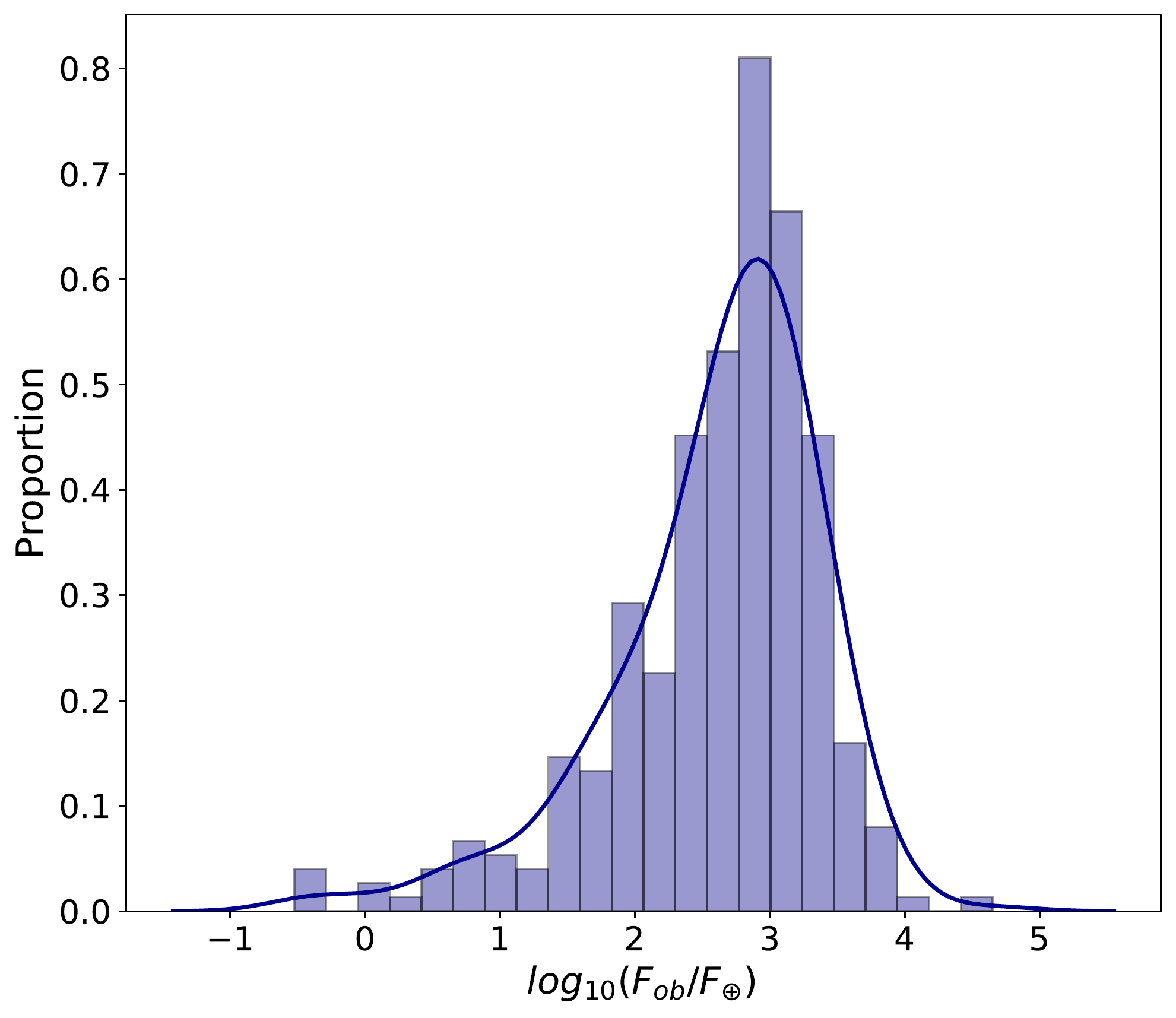}
   \caption{\small
   Histogram of $F_{ob}$ on the logarithmic scale with its kernel density estimation denoted by the blue curve.}
   \label{fig:Fob_marginal}
\end{figure}

Figure \ref{fig:Fob_marginal} shows the empirical histogram (with estimated kernel density overlaid) of the observed flux and visually it appears left-skewed with possibly local modes around the values $1, 2$ and $3$, which indicates that a Gaussian mixture with components having different location parameters can be a reasonable choice. However, notice that we do not need to correctly identify the local modes as those will be estimated from data. Hence we model the the marginal distribution of true flux:

\begin{equation}
    \widetilde{F} \sim \sum_{p=1}^P \pi^{\widetilde{F}}_p \mathcal{N}(\mu^{\widetilde{F}}_p, \sigma^{\widetilde{F}}_p),
    \label{eq:mix_F}
\end{equation}
where $\{\mu^{\widetilde{F}}_p\}_{p=1}^P$, $\{\sigma^{\widetilde{F}}_p\}_{p=1}^P$ are the means and standard deviations of the Gaussian components, and $\{\pi_p\}_{p=1}^P$ are mixture weights. The number of components $P$ will be chosen via model selection methods.

To tackle the label switching problem, order constraints are imposed on the location parameters by a simple reparameterization:

\begin{equation}
    \mu^{\widetilde{F}}_p = \sum_{i=1}^p  a_i, \; p = 1, \dots, P;
    \label{eq:mu_F}
\end{equation}
where $\{a_i\}_{i=2}^P$ are restricted to be positive such that $\mu^{\widetilde{F}}_1, \ldots ,\mu^{\widetilde{F}}_P$ are monotonically increasing.

The priors on the parameters in Equation \ref{eq:mix_F} and  \ref{eq:mu_F} are specified below:

\begin{equation}
\begin{aligned}
    (\pi^{\widetilde{F}}_1, \ldots, \pi^{\widetilde{F}}_P) &\sim \text{Dirichlet}(5, \ldots, 5),\\
    1/\left(\sigma^{\widetilde{F}}_p\right)^2 &\sim \text{Gamma}(0.1, 0.1),\;\mbox{for}\;p = 1, \ldots, P, \\
    a_0 &\sim \mathcal{N}(0, 5),\;\; \mbox{and}\\
    a_p &\sim \mathcal{N}(0, 5)T(0,), \;\mbox{for}\;p = 2, \ldots, P;\\
    \label{eq:F_prior}
\end{aligned}
\end{equation}
where $\text{Gamma}(\alpha, \beta)$ denotes the Gamma distribution with shape $\alpha$ and rate $\beta$, and $T(l,)$ denotes that the distribution is truncated with a lower bound at $l$. 

To make the mixture weights less concentrate on only a few components, the concentration parameters of the Dirichlet prior is larger than one. All other priors are quite vague considering that the largest observed flux after logarithm transformation is around $4.5$. Also note that we place priors on the precision (inverted variance) instead of the standard deviation to be consistent with the parameterization of \textit{JAGS} \citep{Plu03} used for MCMC sampling in this work.

\subsubsection{Conditional Distribution of Radius on Flux}
\label{subsubsec:cond_R}
The insulation flux is one of the many factors that affect the radius. As shown in Figure \ref{fig:mrf_scatter}, the radii of hot Jupiters with high flux are excessively large, and the inflation mechanism seems to be less active for heavier planets. \citet{Dem11, Mil11} find a critical value of flux approximately at 0.2 Gerg $\text{s}^{-1}\text{cm}^{-2}$ below which the inflation mechanism is possibly not significant. Such heterogeneity justifies the use of following FMM for the conditional distribution of radius on flux:

\begin{equation}
    \widetilde{R} | \widetilde{F} \sim \sum_{k=1}^K \pi^{\widetilde{R}}_k \mathcal{N}\left(\mu^{\widetilde{R}}_k(\widetilde{F}), \sigma^{\widetilde{R}}_k\right),
    \label{eq:mix_R}
\end{equation}
where the location parameters $\mu^{\widetilde{R}}_k(\widetilde{F})$ in each component are assumed to be linear in $\widetilde{F}$:

\begin{equation}
    \mu^{\widetilde{R}}_k(\widetilde{F}) = \alpha_k + \beta_k \widetilde{F}, \quad k = 1,\ldots, K.
    \label{eq:mu_R}
\end{equation}

Notice that, this leads to a conditional mean (the so-called regression function) of $\log R$ conditioned on $\widetilde{F}$ given by $E[\widetilde{R} | \widetilde{F}]=\sum_{k=1}^K \pi^{\widetilde{R}}_k(\alpha_k + \beta_k\widetilde{F})$ which is motivated by the top panel in Figure \ref{fig:mrf_scatter}. To prevent the possible label switching problems, we impose order constraints on the precision parameters which allows for identification of the component parameters, and is given by:

\begin{equation}
    1/(\sigma^{\widetilde{R}}_k)^2 = \sum_{i=1}^k  b_i, \; k = 1, \dots, K
    \label{eq:tau_R}
\end{equation}
where $\{b_i\}_{i=1}^K$ are all positive. The above specification leads to a quadratic form of the conditional variance, $\text{Var}(\widetilde{R}|\widetilde{F})$ in terms of $\widetilde{F}$.

The priors on the parameters in Equation \ref{eq:mix_R}, \ref{eq:mu_R} and \ref{eq:tau_R} are listed below:

\begin{equation}
\begin{aligned}
    (\pi^{\widetilde{R}}_1, \ldots, \pi^{\widetilde{R}}_K) &\sim \text{Dirichlet}(5, \ldots, 5), \\
    \alpha_k &\sim \mathcal{N}(0, 5), \\
    \beta_k  &\sim \mathcal{N}(0, 5) \\
    b_k &\sim \text{Gamma}(1,1), \\
    \label{eq:R_prior}
\end{aligned}
\end{equation}
for $k = 1, \ldots, K$.

\subsubsection{Conditional Distribution of Mass on Radius and Flux}

The power law (i.e. $M = CR^\gamma$) has been shown to be effective for characterizing the bivariate mass-radius or radius-period relations on the population level \citep[e.g.][]{Wol16, Wei14, Gie98}. To strengthen the flexibility of fixed values of power law parameters where its coefficients $C$ and $\gamma$ stay unchanged, the broken power law that allows the coefficients to vary across different clusters of planets has been developed \citep[e.g.][]{Ma19, Chen17, Bas17}. 

Note that the observed mass-flux space displays subgroup structures over different ranges of the radius as shown in the bottom panel of Figure \ref{fig:mrf_scatter}. Therefore, we propose an adapted version of the broken power law for the M-R-F Relation:

\begin{equation}
    \widetilde{M}| \widetilde{R}, \widetilde{F} \sim \mathcal{N}(\mu^{\widetilde{M}}(\widetilde{R},\widetilde{F}), \sigma^{\widetilde{M}}(\widetilde{R})),
    \label{eq:mix_M}
\end{equation}
where $\mu^{\widetilde{M}}(\cdot,\cdot)$ is determined by the power law with coefficients varying with $\widetilde{R}$ and is linear in $\widetilde{F}$:

\begin{equation}
    \mu^{\widetilde{M}}(\widetilde{R},\widetilde{F}) = \gamma(\widetilde{R}) + \nu(\widetilde{R}) \widetilde{F}.
    \label{eq:mu_M}
\end{equation}

To capture the subgroup structures, $\sigma^{\widetilde{M}}$, $\gamma$ and $\nu$ are modelled using the zero-th order or linear splines of $\widetilde{R}$ as given below:

\begin{equation}
\begin{aligned}
    1 /\left(\sigma^{\widetilde{M}}(\widetilde{R})\right)^2 &= \sum_{j=1}^J c_j\mathbb{I}(B_{j-1}<\widetilde{R}\leq B_j), \\
    \gamma(\widetilde{R}) &= \sum_{j=1}^J d_j\mathbb{I}(B_{j-1}<\widetilde{R}\leq B_j), \\
    \nu(\widetilde{R}) &= \sum_{j=1}^J \left(g_j + h_j \widetilde{R}\right)\mathbb{I}(B_{j-1}<\widetilde{R}\leq B_j),
    \label{eq:spline_M}
\end{aligned}
\end{equation}
where $-\infty=B_0<B_1<B_2<\cdots<B_{J-1}<B_{J}=\infty$ are the knot points dividing the log-radius dimension into several regions. We use a higher-order (i.e. more smooth) spline to model $\nu(\widetilde{R})$ as it is a key parameter that summarizes the influence of flux on the M-R relation. For the other parameters, we assume them to be a constant within each region to simplify the model. In Section \ref{sec:checking}, we will show that our model is adequate to fit the data.

The spline coefficients $\{g_j, h_j\}_{j=1}^J$ are constrained to keep these functions continuous and avoid abrupt changes the M-R-F relation:
\begin{equation}
g_j + h_jB_j = g_{j+1} + h_{j+1}B_j, \quad j = 1, \ldots, J-1.
\end{equation}

One of the advantages of Bayesian inference is that scientific knowledge about unknown parameters can be utilized for model developments through the priors and still allowing for some level of uncertainty. Both \citet{Ma19} and \citet{Nin18} identify two change points at around $6R_\oplus$ and $11R_\oplus$ in the mass-radius relation. Especially for the inflated hot Jupiters with radii larger than $11R_\oplus$, a flat mass-radius relation is observed possibly due to an unknown heating mechanism related the level of insulation flux \citep{Gui02,kov10, Lau11}. 

Therefore, we choose $J=3$ in this work and place informative priors on the knots $B_1$ and $B_2$:

\begin{equation}
\begin{aligned}
    B_1 &\sim \mathcal{N}(\log_{10} 6, 0.2),\\
    B_2 &\sim \mathcal{N}(\log_{10} 11, 0.2)T(B_1, ).\\
    \label{eq:prior_B}
\end{aligned}
\end{equation}

The priors on the rest of the parameters are listed below:

\begin{equation}
\begin{aligned}
    c_j &\sim \text{Gamma}(0.1, 0.1)\\
    d_j &\sim \mathcal{N}(0, 2) \\
    h_j &\sim \mathcal{N}(0, 1) \\
    \label{eq:prior_M}
\end{aligned}
\end{equation}

for $j = 1, \ldots, J$, and $g_1 \sim \mathcal{N}(0, 2)$. It is to be noted, although we center the knots at values approximately identified by previous literature, we still allow for uncertainty around these values and let data decide on the estimated values.

\subsubsection{Model for Measurement Errors}
The observations $M^{ob}$, $R^{ob}$ and $F^{ob}$ are subject to the (known) measurement errors $\sigma^{Mob}$,  $\sigma^{Rob}$ and  $\sigma^{Fob}$ that usually depends on experimental conditions such as the instruments and different mechanisms used to observe the mass, radius and flux of the planets. We follow the treatment in \citet{Wol16, Nin18, Ses18, Ma19} that assumes that the observed values of the mass, radius and flux are normally distributed around the unobserved values $M$, $R$ and $F$ of the corresponding mass, radius and flux, respectively and use the following measurement error model:

\begin{equation}
\begin{aligned}
    M^{ob} &\sim \mathcal{N}(M, \sigma^{Mob})\\
    R^{ob} &\sim \mathcal{N}(R, \sigma^{Rob})\\
    F^{ob} &\sim \mathcal{N}(F, \sigma^{Fob})\\
    \label{eq:observe}
\end{aligned}
\end{equation}
The above normality assumption is mostly driven by convenience rather than physics, but, unfortunately this topic has received a very little attention in astronomy literature. We provide a way to validate its rationality in section \ref{subsec:residual}. 

\subsection{Model Selection}
All of the above FMMs developed in earlier sections requires the specification of the number of components. There are two such tuning parameters in our model: $P$ and $K$ that are the number of components in the mixtures. With larger values of $P$ and $K$, our model built on Gaussian mixtures could result in multimodal posterior distributions that may not be efficiently explored by MCMC samplers even with order restrictions on the mean or standard deviation parameters. The primary reason for this is due to the fact with larger components, it becomes very unlikely for MCMC samplers to visit the components with very low weights. So, we restrict the choice of number of components to lower values and iteratively use convergence diagnostics of MCMC and standard model selection criteria to select such tuning model parameters. 

We also need to make judicious choices for the number of knots used for the broken power law for our model. \citet{Bas17} and \citet{Buc14} find transitional points in the mass-radius relation at $3.9R_\oplus$ and $12.1R_\oplus$ respectively, and the additional knots have informative priors centered around them.

Therefore, we first assess the convergence of the candidate parameterizations with different sets of tuning parameters and knots. The Gelman-Rubin (GR) diagnostic $\hat{R}$ \citep{Gel92} that measures the discrepancies between parallel Markov chains is calculated, and $\hat{R} < 1.1$ usually indicates convergence. 

For the models appears to converge by the $\hat{R}$ criteria and lead to unimodal posteriors, we further calculate their deviance information criterion (DIC) \citep{Spi02} that measures the goodness of fit and penalizes model complexity. A model with smaller DIC is preferred, and the difference between DICs indicates the degree of the preference.

MCMC sampling in this work is performed by JAGS that is also capable of calculating DIC of a hierarchical model directly \citep{Plu03}. It turns out that only the two-knot configuration specified in Equation \ref{eq:prior_B} leads to converging unimodal posteriors. We also find that the smallest DIC is achieved at $P=2$ and $K=3$ although only by a very narrow margin. 
It's important to note that DIC may not be suited to select mixture models \citep{Ste10}. In fact, since the Gaussian mixtures serve as internal parts of our model, accurate selection of the tuning parameters $P$ and $K$ are not that influential as long as the model is adequate for approximating the underlying M-R-F relation. Thus, we continue to choose $P=2$ and $K=3$ in our final model configuration and perform model checking to validate its adequacy in section \ref{sec:checking}. 

\section{Results} \label{sec:results}

We ran 4 parallel chains, each consisting of 400,000 iterations. The first 200,000 iterations were discarded as burn-ins to allow for reasonable mixing, and the remaining 200,000 iterations were thinned with a lag of 20 (for computational efficiency). The GR diagnostic for each parameter calculated from the combined 40,000 posterior samples was considerably less than 1.1, which indicates that the chains have mixed well and there are no apparent issues with MCMC convergence.

The posterior estimates of the parameters of the marginal density $f(\widetilde{F})$ is shown in Table \ref{tab:pos_F}. As the weights shows the abundance of samples in both components, the posterior estimates are all well constrained.

\begin{table}
\centering
\caption{Posterior mean of the parameters in $f(\widetilde{F})$ with $P=2$. The error bars correspond to the 16th and 84th posterior percentiles.}
\label{tab:pos_F}
\begin{tabular}{cccc}
\hline
Comp. ID & Weight & Mean & S.D.\\
p & $\pi^{\widetilde{F}}_p$ & $\mu^{\widetilde{F}}_p$ & $\sigma^{\widetilde{F}}_p$\\
\hline
1 & $0.36^{+0.07}_{-0.07}$ & $2.00_{-0.16}^{+0.16}$ & $0.94_{-0.08}^{+0.08}$\\
2 & $0.64^{+0.07}_{-0.07}$ & $2.91_{-0.04}^{+0.04}$ & $0.39_{-0.04}^{+0.04}$\\
\hline
\end{tabular}
\end{table}

Table \ref{tab:pos_R} lists the posterior estimates of the parameters in $f(\widetilde{R}|\widetilde{F})$. Since $\beta_1$ is around $0$, the planets belonging to the first component show weak correlation between radius and flux. It is observed in the other two components that the radius inflates with increasing flux, while the second component has higher heating efficiency ($\beta_2 = 0.35$ vs $\beta_3 = 0.10$). Also note that the estimates for the first component are not as well constrained as others due to the lack of samples ($w^{\widetilde{R}}_1 = 0.1$).

\begin{table}
\centering
\caption{Posterior mean of the parameters in $f(\widetilde{R}|\widetilde{F})$ with $K=3$. The error bars correspond to the 16th and 84th posterior percentiles.}
\label{tab:pos_R}
\begin{tabular}{cccc}
\hline
Comp. ID & Weight & Mean & S.D.\\
k & $\pi^{\widetilde{R}}_k$ & $\alpha_k+\beta_k \widetilde{F}$ & $\sigma^{\widetilde{R}}_k$\\
\hline
1 & $0.10^{+0.02}_{-0.02}$ & $0.29_{-0.10}^{+0.10}+0.00_{-0.04}^{+0.04} \widetilde{F}$ & $0.15_{-0.02}^{+0.02}$  \\
2 & $0.24^{+0.05}_{-0.05}$ & $0.07_{-0.09}^{+0.09}+0.35_{-0.03}^{+0.03} \widetilde{F}$ & $0.12_{-0.01}^{+0.01}$  \\
3 & $0.66^{+0.05}_{-0.05}$ & $0.85_{-0.03}^{+0.03}+0.10_{-0.01}^{+0.01} \widetilde{F}$ & $0.12_{-0.01}^{+0.01}$  \\
\hline
\end{tabular}
\end{table}

\begin{figure*}
    \centering
    \begin{subfigure}{0.5\textwidth}
      \centering
      \includegraphics[width=8.5cm]{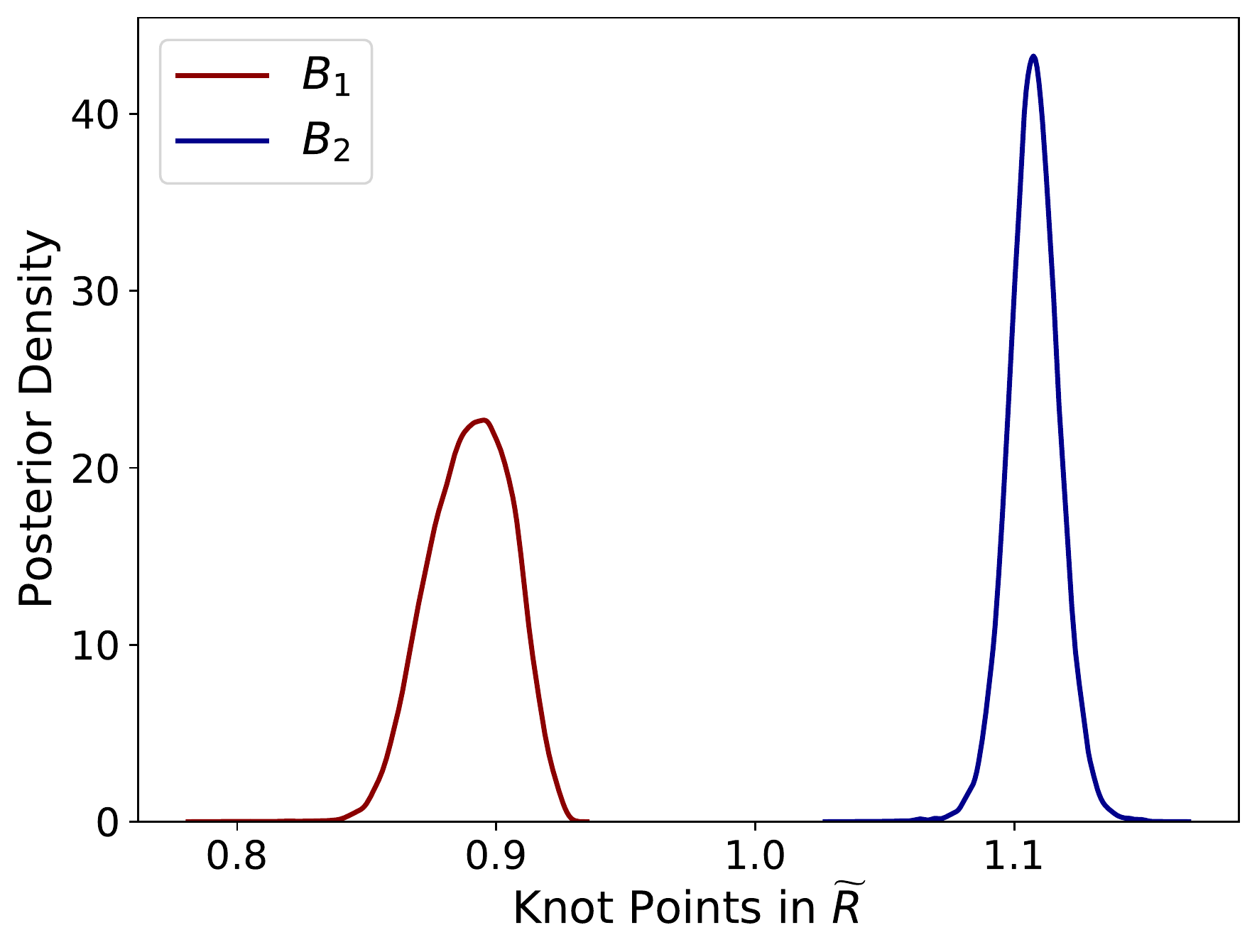}
      \caption{}
    \end{subfigure}%
    \begin{subfigure}{0.5\textwidth}
      \centering
      \includegraphics[width=8.5cm]{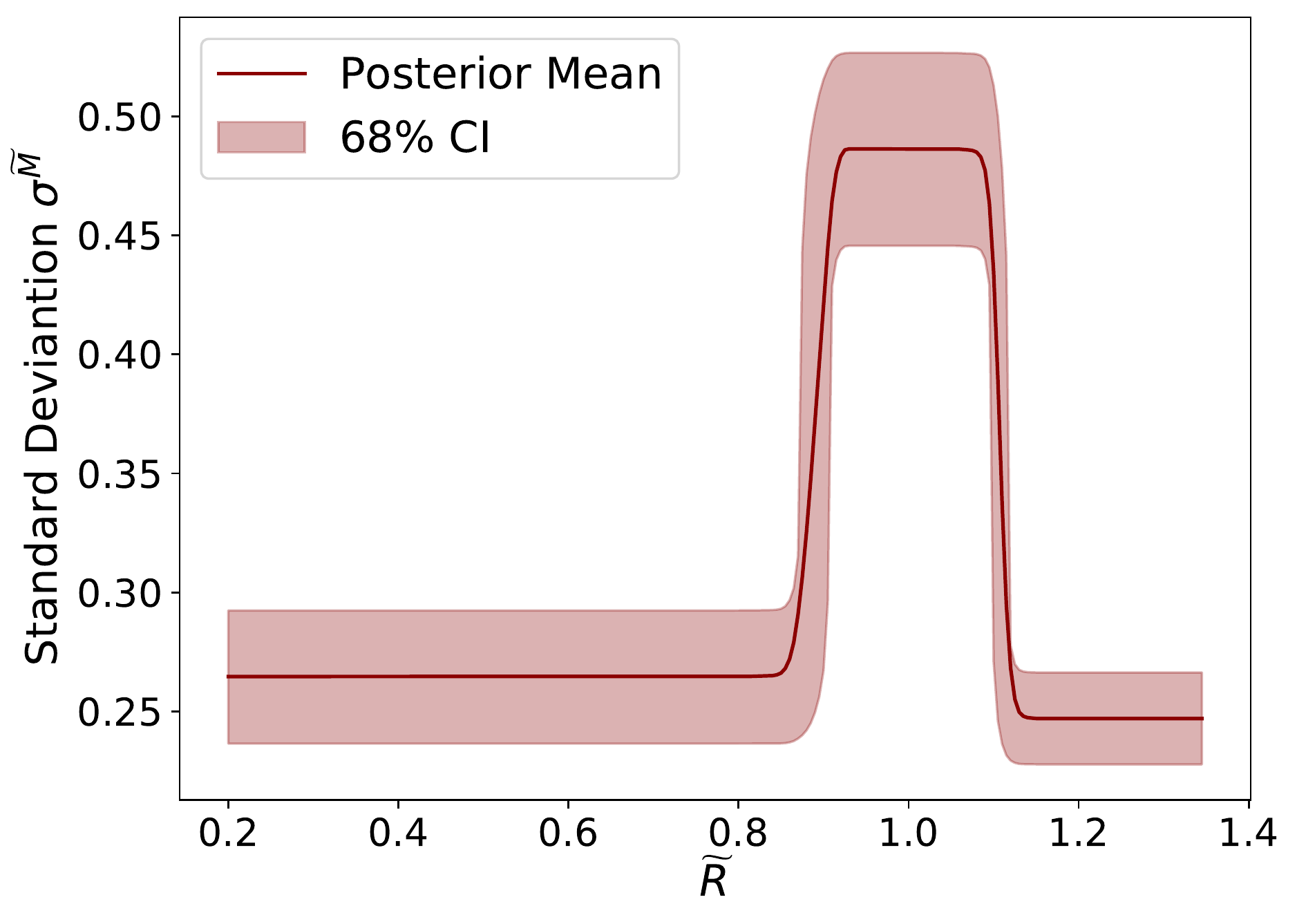}
      \caption{}
    \end{subfigure}
    
    \begin{subfigure}{0.5\textwidth}
      \centering
      \includegraphics[width=8.5cm]{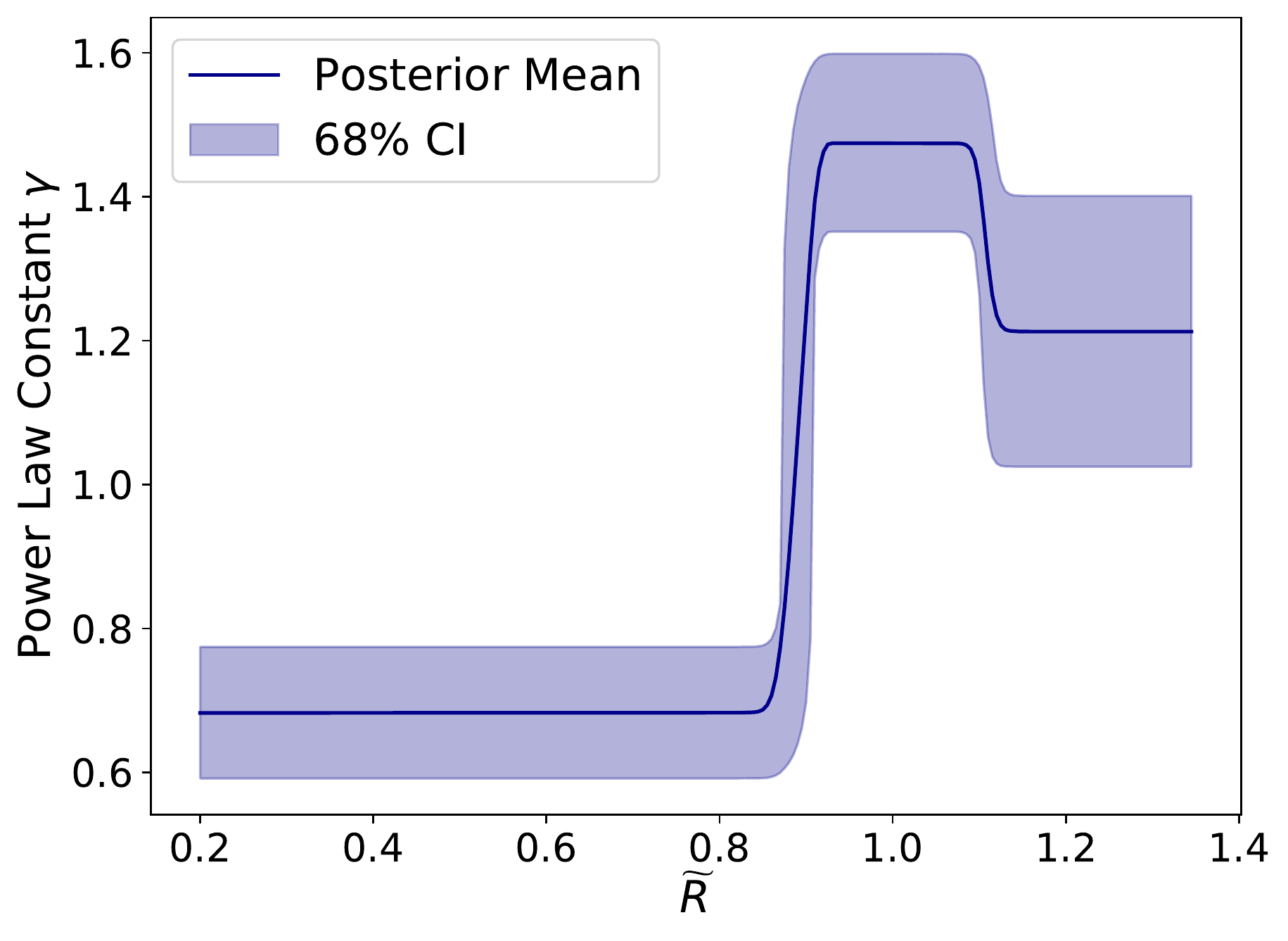}
      \caption{}
    \end{subfigure}%
    \begin{subfigure}{0.5\textwidth}
      \centering
      \includegraphics[width=8.5cm]{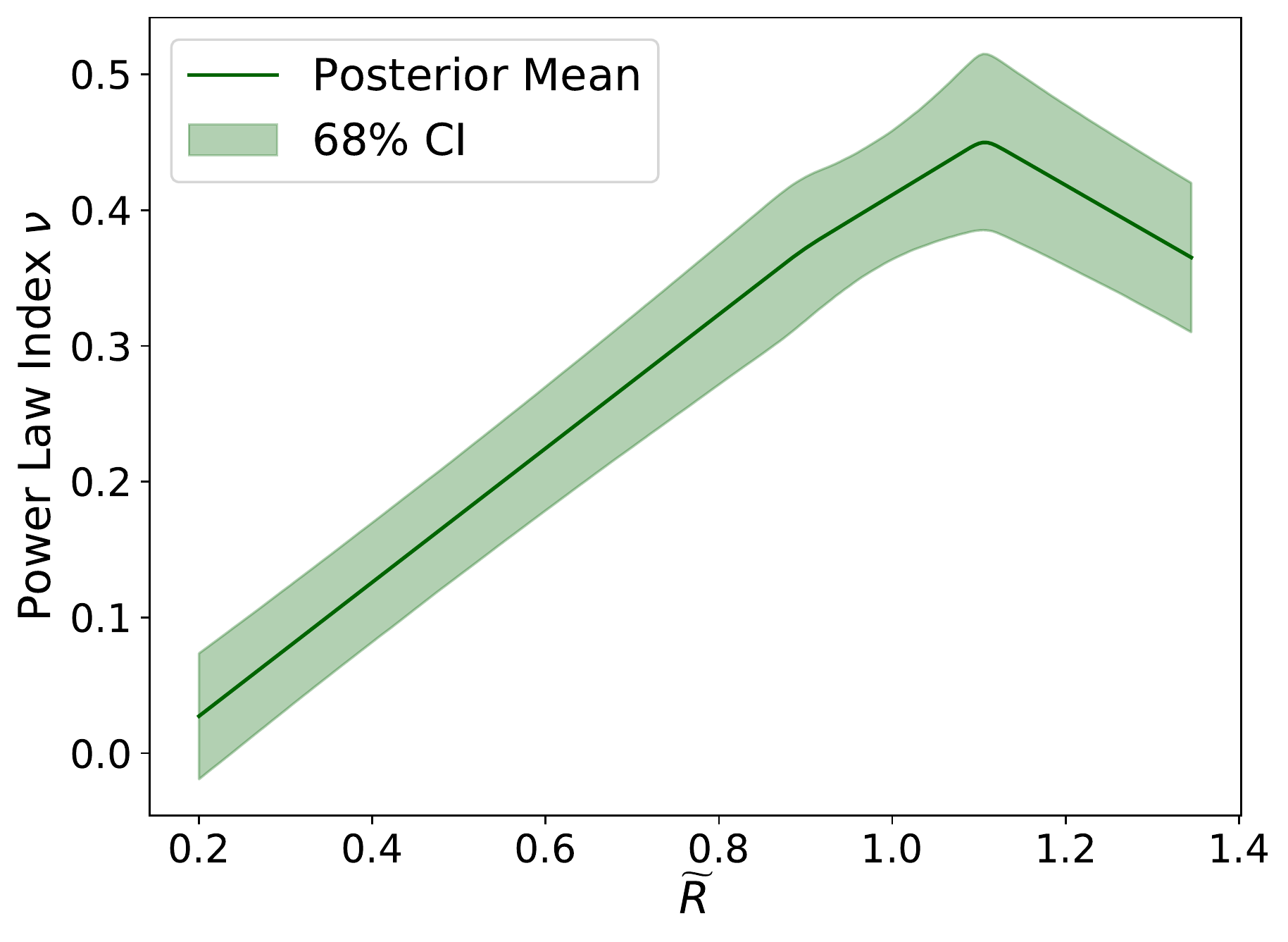}
      \caption{}
    \end{subfigure}
    
    \begin{subfigure}{0.5\textwidth}
      \centering
      \includegraphics[width=8.5cm]{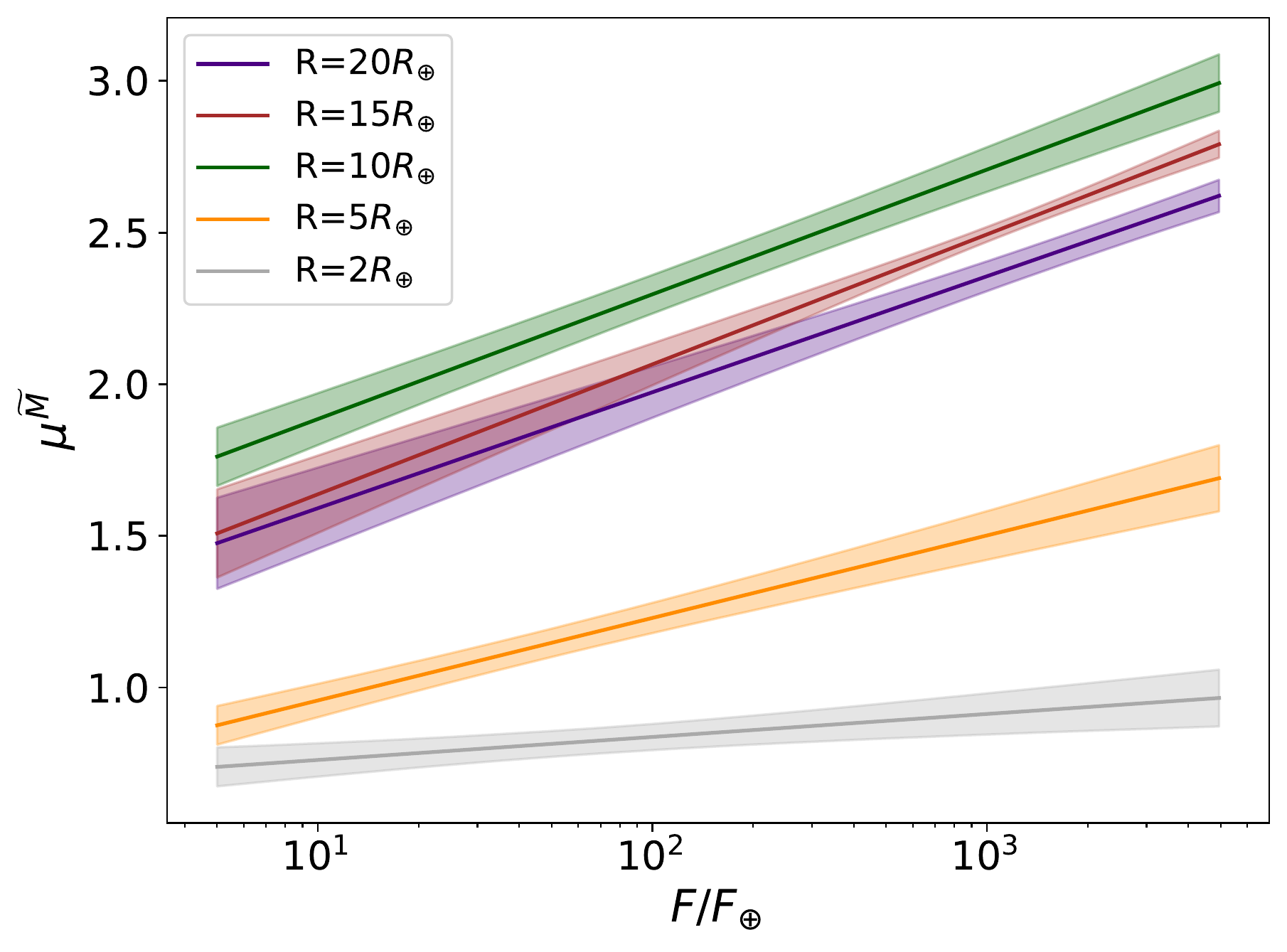}
      \caption{}
    \end{subfigure}%
    \begin{subfigure}{0.5\textwidth}
      \centering
      \includegraphics[width=8.5cm]{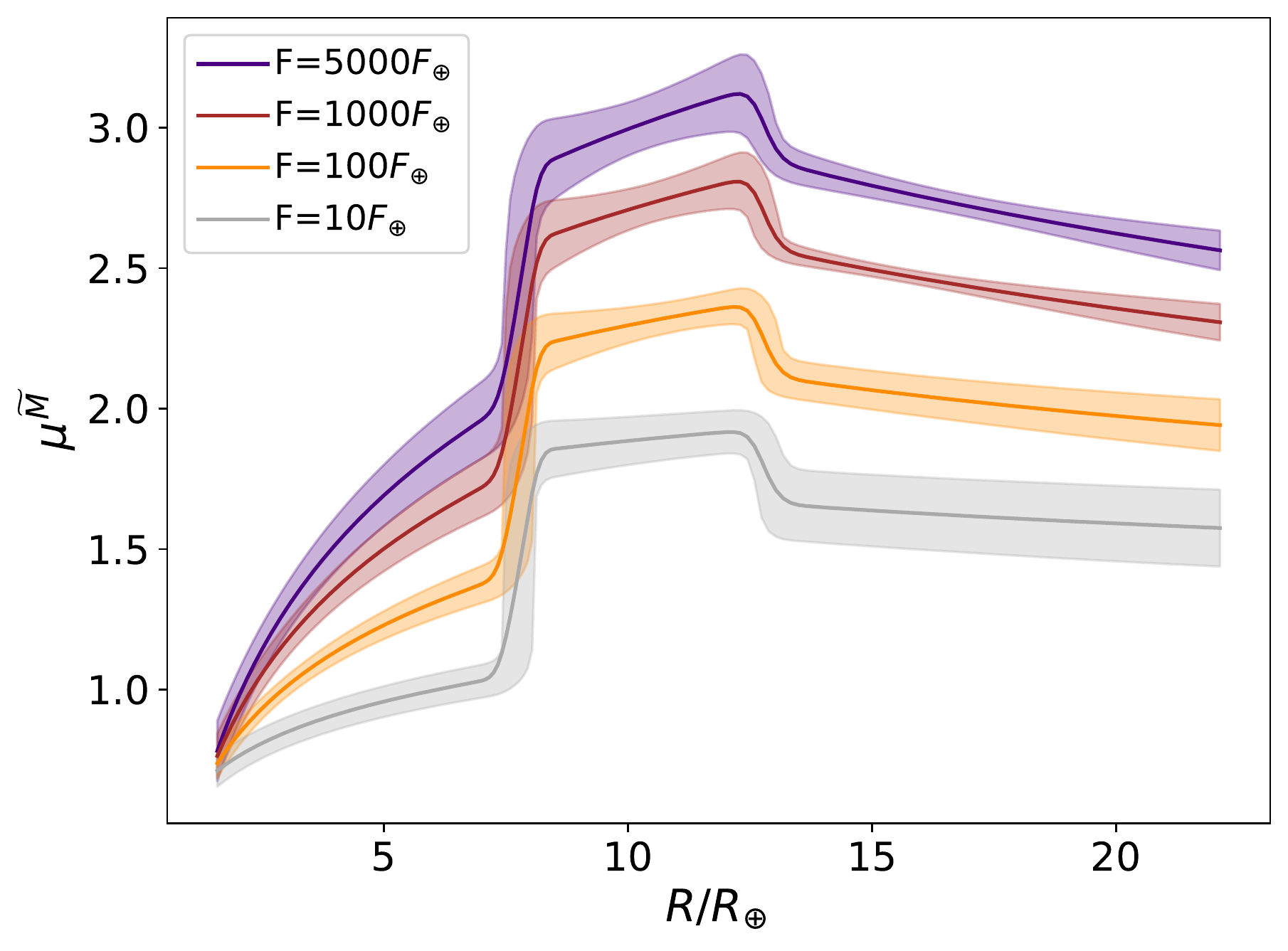}
      \caption{}
    \end{subfigure}
    
  \caption {(a): Posterior distribution of knot points $B_1$ and $B_2$. (b), (c) and (d): The posterior means and the 68\% credible intervals (CI) of $\sigma^{\widetilde{M}}$, power law constant $\gamma$ and index $\nu$ in terms of $\widetilde{R}$ (logarithmic scale). (e) and (f): The posterior mean and the 68\% credible interval of $\mu^{\widetilde{M}}$ in terms of $F$ and $R$ (linear scale) respectively. \label{fig:M_pos}}
\end{figure*}

Figure \ref{fig:M_pos}(a) shows the posteriors of the transitional points $B_1$ and $B_2$ where $f(\widetilde{M}|\widetilde{R}, \widetilde{F})$ modelled by the broken power law exhibits significant changes. With $B_1$ and $B_2$ having posterior estimates at $0.89_{-0.02}^{+0.02}$ and $1.11_{-0.01}^{+0.01}$ on logarithmic scale respectively, the M-R-F relation is divided into three regions roughly corresponding to Neptunes, Jupiters, and super-Jupiters. By including the effect of flux, the transitional points from our model are both larger than those obtained by by \citet{Ma19, Nin18}.

The posterior estimates of the broken power law coefficients as functions of the radius are displayed in Figure \ref{fig:M_pos}(c)(d). In the first two regions where radius is less than around $13R_\oplus$, the broken power law constant $\gamma$ and index $\nu$ increase with larger radius. However, an opposite trend is observed for super-Jupiters. The intrinsic scatter $\sigma^{\widetilde{M}}$ also has its highest value around $12R_\oplus$ as shown in Figure \ref{fig:M_pos}(b).

To better understand the M-R-F relation estimated by our model, Figure \ref{fig:M_pos}(e) shows $\mu^{\tilde{M}}$ as a function of the flux. In general, it states that the planetary mass increases with higher level of flux with Jupiters having the largest increasing rate.

We also plot the estimated M-R-F relation as a function of radius at different flux levels in Figure \ref{fig:M_pos}(f) where several patterns are observed. First, the planets receiving higher level of flux are denser. It could be attributed to the stronger evaporation of H/He envelope triggered by higher stellar XUV flux \citep[e.g.][]{Lam03,Yel04,Bar04}, which leads to a larger fraction of heavy elements. For low-mass and high-flux planets, they could have no (or very thin) H/He envelope, only rock/iron cores \citep{Lop12}. Second, planetary mass and radius have positive correlation in the first two radius regions, which is as expected and in agreement with the literature on the mass-radius relation \citep[e.g.][]{Wol16, Chen17, Nin18, Ma19}. Finally, the estimated mean log-mass drops as the radius becomes larger after around $13R_\oplus$ at high flux (the blue and red curves). For highly inflated hot-Jupiters, the H/He envelope dominates the composition and is only loosely attached to the planetary core. Thus, significant mass loss by stellar irradiation may occur during their evolution \citep{Val10, Hub07}. The model of \citet{Bar04} also suggests that for close-in giant planets with high flux and at a late evolutionary stage, the radius could increase rapidly as the outer layer expands violently, which further amplifies the atmospheric mass loss rate.

\section{Model Checking} \label{sec:checking}
In this section, we provide two novel approaches for model checking that aim to validate the consistency of our hierarchical model with the observed data, as well as identify possible directions for model improvements.

\subsection{Robust Hotelling's Test}\label{subsec:hotelling}

A popular tool for Bayesian model checking is the posterior predictive p-value (PPP) that measures the discrepancy between the simulated data from the posterior predictive distribution and the actual data that have been observed. PPP is easy to implement with posterior samples, but requires test statistics that manages to summarize the model behaviors. \citet{Wol16} propose two test statistics to check their hierarchical model on the mass-radius relation, which are also adopted by \citet{Ses18}. However, those test statistics only focus on one level of the hierarchical model and thus fail to assess the model's ability to characterize the joint distribution. Therefore, we propose an approach based on the robust Hotelling's test to test the model adequacy by checking if the mean of the posterior joint distribution of $(M_i, R_i, F_i)$ estimated by the model agrees with the observed values $(M_i^{ob}, R_i^{ob}, F_i^{ob})$. 

Hotelling's test \citep{Hot31} is the multivariate counterpart of the well-known t-test. Let $\{\boldsymbol{x}_1, \ldots, \boldsymbol{x}_n\} \in \mathbf{R}^p$ be a random sample from a p-variate normal distribution with location $\boldsymbol{\mu}$ and covariance $\mathbf{\Sigma }$. Under the null hypothesis $H_0: \boldsymbol{\mu} = \boldsymbol{\mu}_0$, the Hotelling’s $T^2$ statistic follows a scaled F distribution:

\begin{equation}
    T^2 \equiv n(\overline{\mathbf{x}}-\boldsymbol{\mu}_0)^T\mathbf{S}^{-1}(\overline{\mathbf{x} }-\boldsymbol{\mu}_0) \sim \frac{(n-1)p}{n-p}F_{p,n-p}
\end{equation}

where $\overline{\mathbf{x}} =\frac{1}{n}\sum_{i=1}^{n} \mathbf{x}_i$ is the sample mean, $\mathbf{S} = \frac{1}{n-1}\sum_{i=1}^n(\mathbf{x}_{i}-\overline{\mathbf{x}})(\mathbf{x}_{i}-\overline{\mathbf {x}})^T$ is sample covariance, and $F_{p,n-p}$ denotes the F distribution with degrees of freedom $p$ and $n-p$.

The simultaneous confidence intervals for each component of $\boldsymbol{\mu}$ at the significant level of $100(1-\alpha)\%$ are given by:

\begin{equation}
    \left(\overline{\mathbf{x}}_i \pm \sqrt{\frac{p(n-1)}{n(n-p)}F_{p,n-p}(\alpha)s_{ii}}\right)
\end{equation}

where $\overline{\mathbf{x}}_i$ is the $i$th entry of $\overline{\mathbf{x}}$, $s_{ii}$ is the $i$th diagonal entry of $\mathbf{S}$, and $F_{p,n-p}(\alpha)$ is the upper $\alpha$th quantile of $F_{p,n-p}$.

The classic Hotelling's test has several optimality properties including the robustness to moderate departures from normality \citep[e.g.][]{Mar75, Kar81, Dem06}. More recent works on Hotelling's test \citep[e.g.][]{Wil02, Van13} seek to improve its robustness to outliers by replacing the naive location and covariance estimators (i.e. $\overline{\mathbf{x}}$ and $\mathbf{S}$) with their robust counterparts.

We justify our model by checking if $(M_i, R_i, F_i)$ is close to the mean of the posterior samples $\{(M_i^{(s)}, R_i^{(s)}, F_i^{(s)})\}_{s=1}^S$, where $S$ is the number of posterior samples. To address the problem that only the measurements of $(M_i, R_i, F_i)$ with uncertainties are available, we perform the test using following steps:

\begin{enumerate}
    \item Define the $1\sigma$ "observed" hypercube:
\begin{equation}
   O_i = (M^{ob}_i \pm \sigma^{Mob}_i, R^{ob}_i \pm \sigma^{Rob}_i, F^{ob}_i \pm \sigma^{Fob}_i)
\end{equation}

    \item Compute the simultaneous 68\% confidence intervals of the mean of $\{(M_i^{(s)}, R_i^{(s)}, F_i^{(s)})\}_{s=1}^S$, which is also represented by a hypercube
    
\begin{equation}
   C_i = ((M^l_i, M^u_i), (R^l_i, R^u_i), (R^l_i, R^u_i))
\end{equation}
    
    where the superscripts $l$ and $u$ denotes the lower and upper bounds respectively.
    
    \item Check if $O_i$ and $C_i$ intersect.
\end{enumerate}

The motivation of the first step is that the true values should be close to the observed ones, such that $(M_i, R_i, F_i)$ resides within $O_i$ with a high probability. The second step adopts the robust Hotelling test developed by \citet{Wil02} and implemented in the R package {\tt rrcov}.

The preceding procedure is repeated for each of the 319 planets in our sample. The result shows that $C_i$ is fully contained in $O_i$ for 310 planets, indicating that the posterior joint distribution of $(M_i,R_i,F_i)$ concentrates closely around the true values for the majority of the sample planets. 

There are 4 planets whose $C_i$ do not intersect with $O_i$, including HATS-61 b, Kepler-87 b, WASP-140 b and WTS-1 b. They can be treated as outliers with respect to our model. Among the sample planets, HATS-61 b, WTS-1 b and WASP-140 b have significantly larger mass than the others with similar radii and fluxes. On the other hand, the flux received by Kepler-87 b is much lower than the others with similar radii and masses.

\subsection{Bayesian Studentized Residual}\label{subsec:residual}

Residual analysis is a common tool for detecting outlying data points and validating normal assumption on the error terms in frequentist linear regression models. Let the regression model be $Y_i = \boldsymbol{x}_i^T \boldsymbol{\beta} + \epsilon_i$ for $i=1,\ldots,n$, where $\boldsymbol{\beta} \in \mathbb{R}^p$ is the regression coefficients, $\boldsymbol{X} = (\boldsymbol{x}_1, \ldots, \boldsymbol{x}_n)^T$ is the design matrix, $\epsilon_i$ are i.i.d. error terms from $\mathcal{N}(0, \sigma)$. The simple additive residual is defined as $e_i = y_i - \boldsymbol{x}_i^T \widehat{\boldsymbol{\beta}}$, where $\widehat{\boldsymbol{\beta}}$ is the ordinary least square estimate of $\boldsymbol{\beta}$. To deal with the problem that $e_i$ scales with the magnitude of $Y_i$, the (internally) studentized residual are given by:

\begin{equation}
    r_i = \frac{e_i}{s(e_i)} = \frac{e_i}{\widehat{\sigma}\sqrt{1-h_{ii}}}
\label{eq:residual}
\end{equation}
where $s(e_i)$ is the estimated of the standard deviation of $e_i$, $\widehat{\sigma}$ is the estimate of $\sigma$ and is usually given by $\widehat{\sigma} = \sqrt{\sum_{j=1}^n e_j^2 / (n-p)}$ in classical linear models, and the leverage $h_{ii}$ is the $i$th diagonal entry of the projection or hat matrix $H = \boldsymbol{X}(\boldsymbol{X}^T \boldsymbol{X})^{-1}\boldsymbol{X}^T$.

For a data set where $n$ is much larger than $p$, $r_i$ approximately follows a standard normal distribution. Therefore, an observation with $|r_i|$ larger than 3 can be treated as a outlier. And if $\{r_i\}_{i=1}^n$ don't seem to arise from a normal distribution, the assumption of normality should be further investigated.

The normality assumption in many Bayesian hierarchical models often corresponds to a linear model. For example, Equation \ref{eq:mix_M} and \ref{eq:mu_M} can be rewritten as

\begin{equation}
    \widetilde{M}_i = d + g\widetilde{F}_i + h\widetilde{F}_i \cdot\widetilde{R}_i + \epsilon_{i} \quad \epsilon_{i} \sim \mathcal{N}(0, \frac{1}{\sqrt{c}}) \;\;\; i = 1,\ldots,n
\label{eq:linear1}
\end{equation}

where the region index for the coefficients are dropped for simplicity.

Therefore, given a posterior sample of model parameters $\Theta^{(s)}$, we can calculate its studentized residual denoted by $\{\widetilde{r}_i^{(s)}, s=1,\ldots,S\}$ following Equation \ref{eq:residual}, where $S$ is the number of posterior samples. Then the Bayesian studentized residual $\widetilde{r}_i$ is defined as the average of them \footnote{With the broken power law applied, there are in fact $J$ linear models. A planet's studentized residual is computed using the coefficients of the radius region where it belongs.}.

\begin{figure}
\centering
   \includegraphics[width=8.5cm]{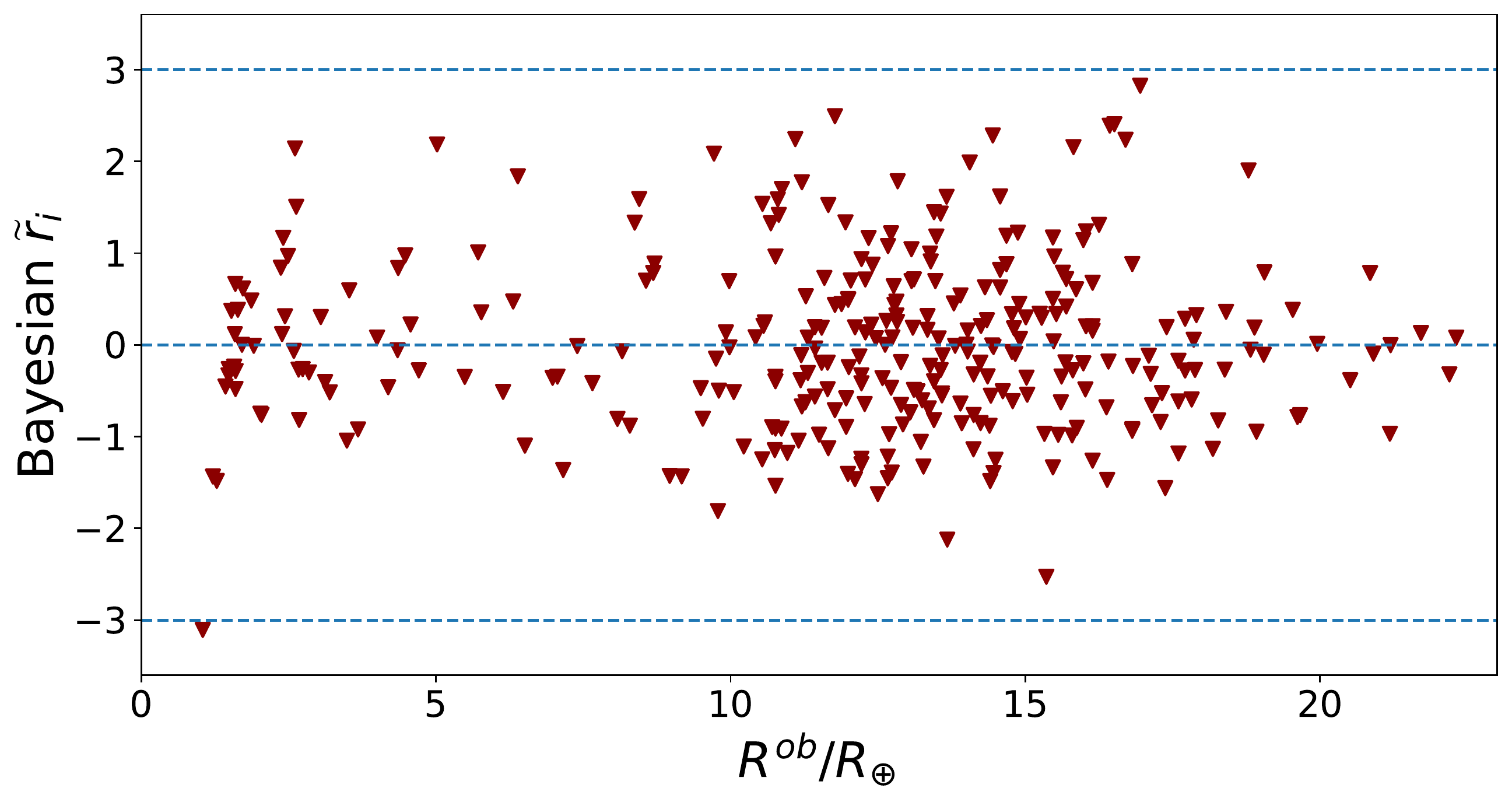}
   \caption{\small
   Posterior mean of $r_i$ \textit{vs} observed radii.}
   \label{fig:residual_logM}
\end{figure}

In Figure \ref{fig:residual_logM}, we plot the Bayesian studentized residuals for each sample planet, which shows that the majority of studentized residuals scatter between $-3$ and $+3$. The only potential outlying planet is the low-mass and earth-sized TRAPPIST-1f orbiting around an ultracool dwarfs. Due to the lack of similar planets in the sample, our model has difficulty in characterizing its M-R-F relation.

Similar residual analysis can be applied to the observational layer (see Equation \ref{eq:observe}) that corresponds to a very simple linear model
\begin{equation}
    M^{ob}_i = M + \epsilon_{i} \quad \epsilon_{i} \sim \mathcal{N}(0, \sigma^{Mob}_i) \;\;\; i = 1,\ldots,n
\end{equation}

\begin{figure}
\centering
   \includegraphics[width=8.5cm]{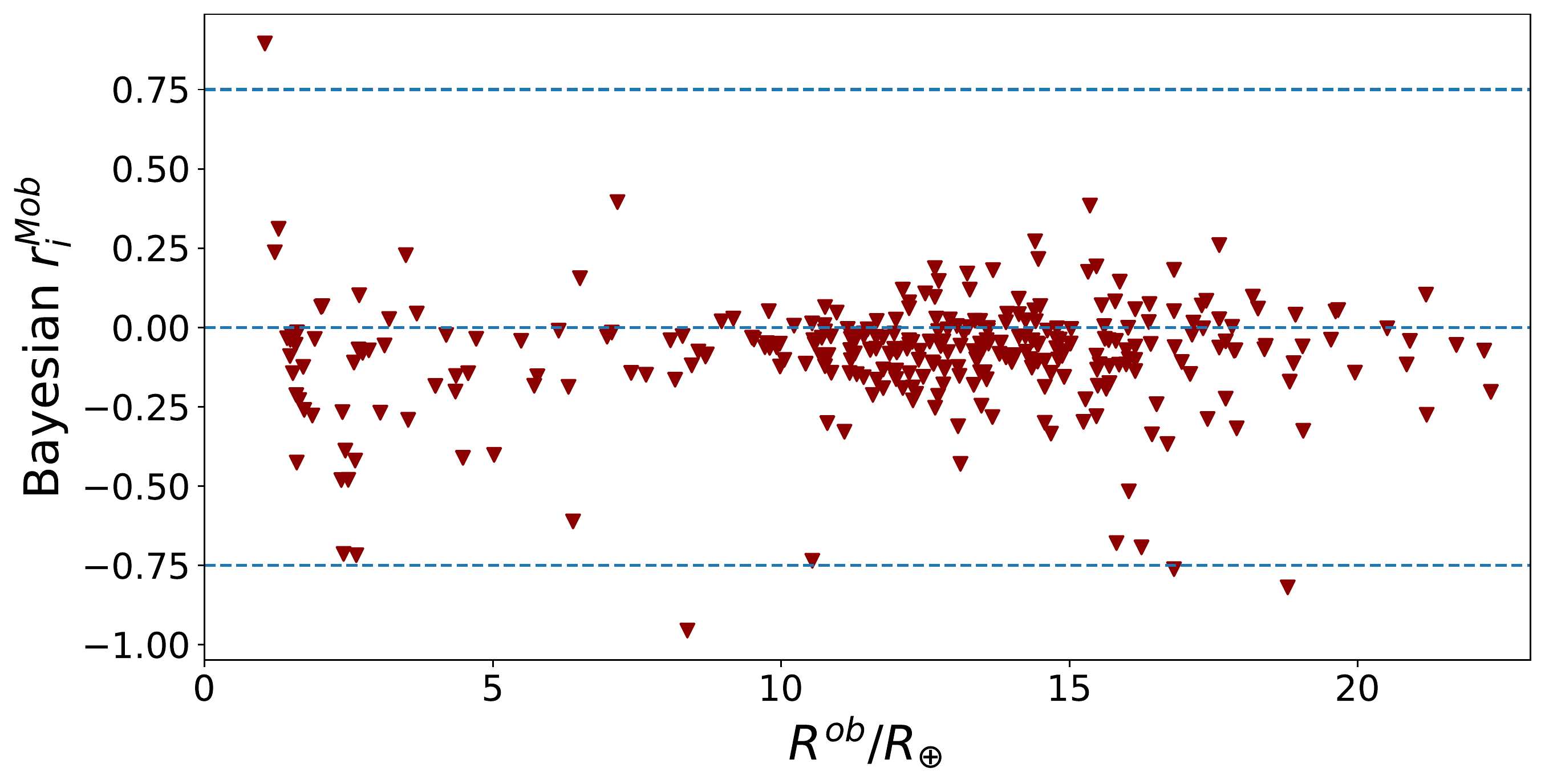}
   \caption{\small
   Posterior mean of $r_i^{Mob}$ \textit{vs} observed radii.}
   \label{fig:residual_Mob}
\end{figure}

\begin{figure*}
\centering
\begin{minipage}[b]{.5\textwidth}
  \centering
  \includegraphics[width=.85\linewidth]{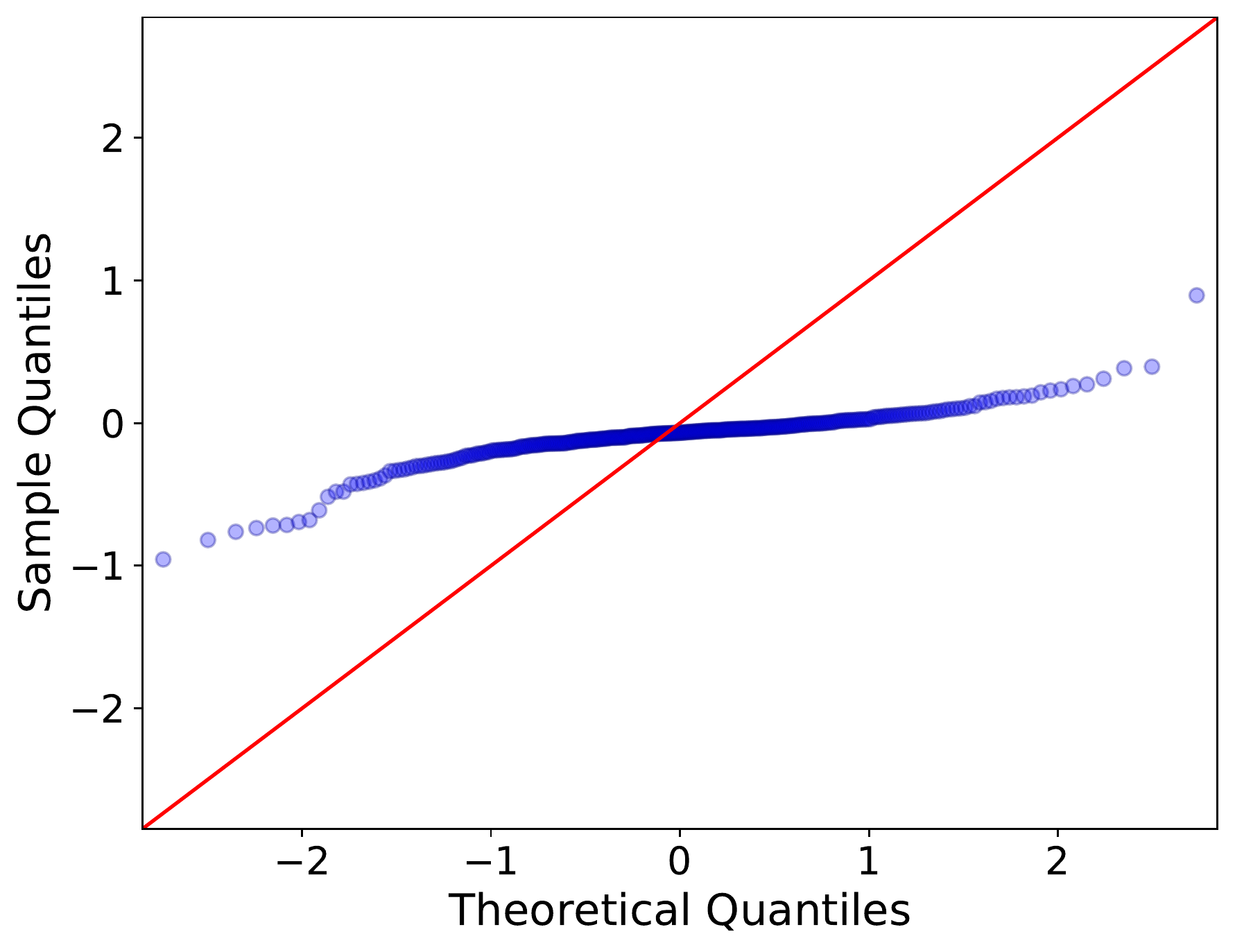}
\end{minipage}%
\begin{minipage}[b]{.5\textwidth}
  \centering
  \includegraphics[width=.85\linewidth]{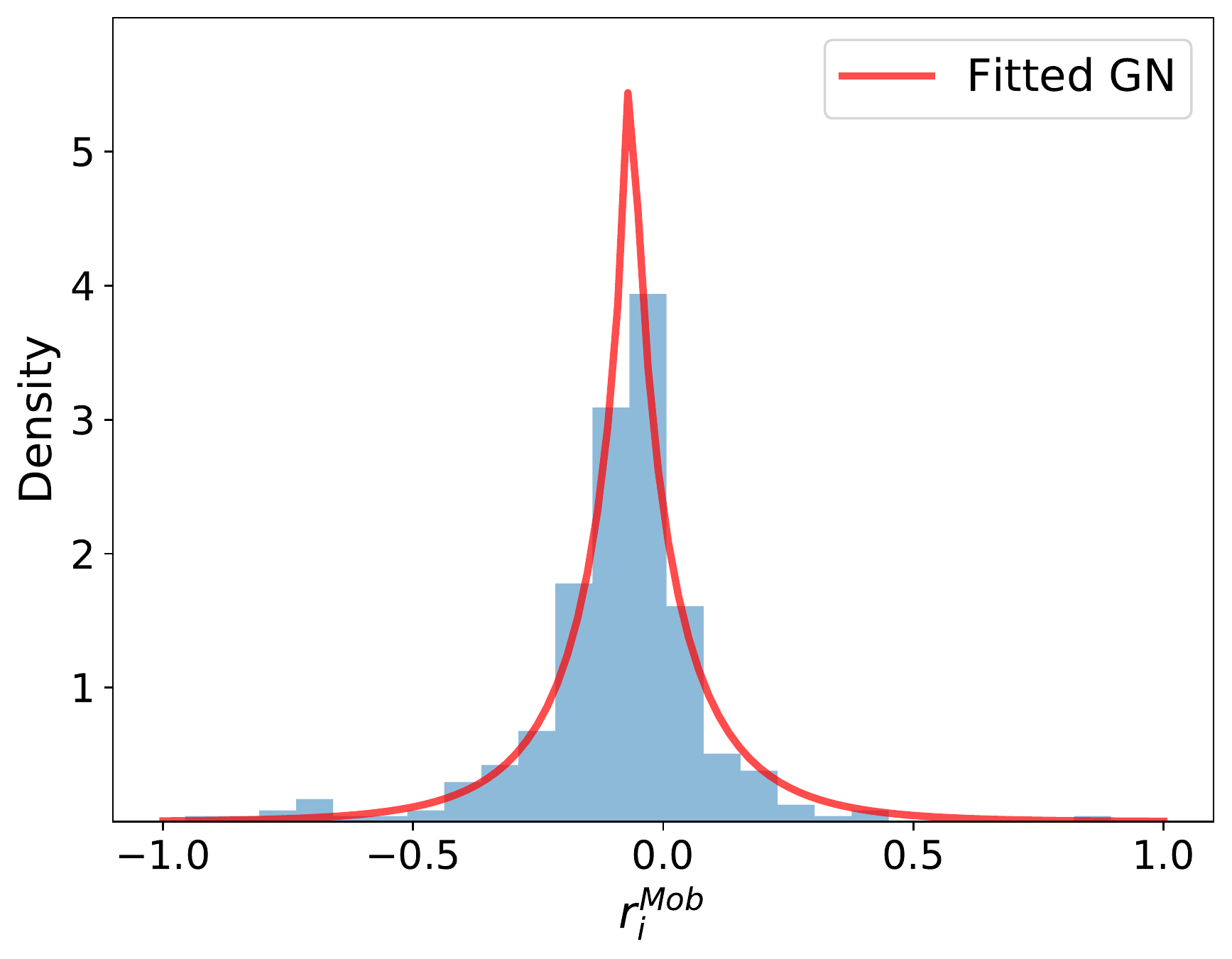}
\end{minipage}
\par
\begin{minipage}[t]{.5\textwidth}
  \centering
  \caption{Q-Q plot of $r_i^{Mob}$.}  
  \label{fig:QQ_residual_Mob}  
\end{minipage}%
\begin{minipage}[t]{.5\textwidth} 
  \centering
  \caption{Histogram and empirical fitting curve of $r_i^{Mob}$ using $\text{GN}(-0.06, 0.06, 0.73)$. The fitted parameters are obtained by maximum likelihood estimation.}  
  \label{fig:GN_residual_Mob}  
\end{minipage} 
\begin{minipage}{.5\textwidth}
  \centering
  \includegraphics[width=.85\linewidth]{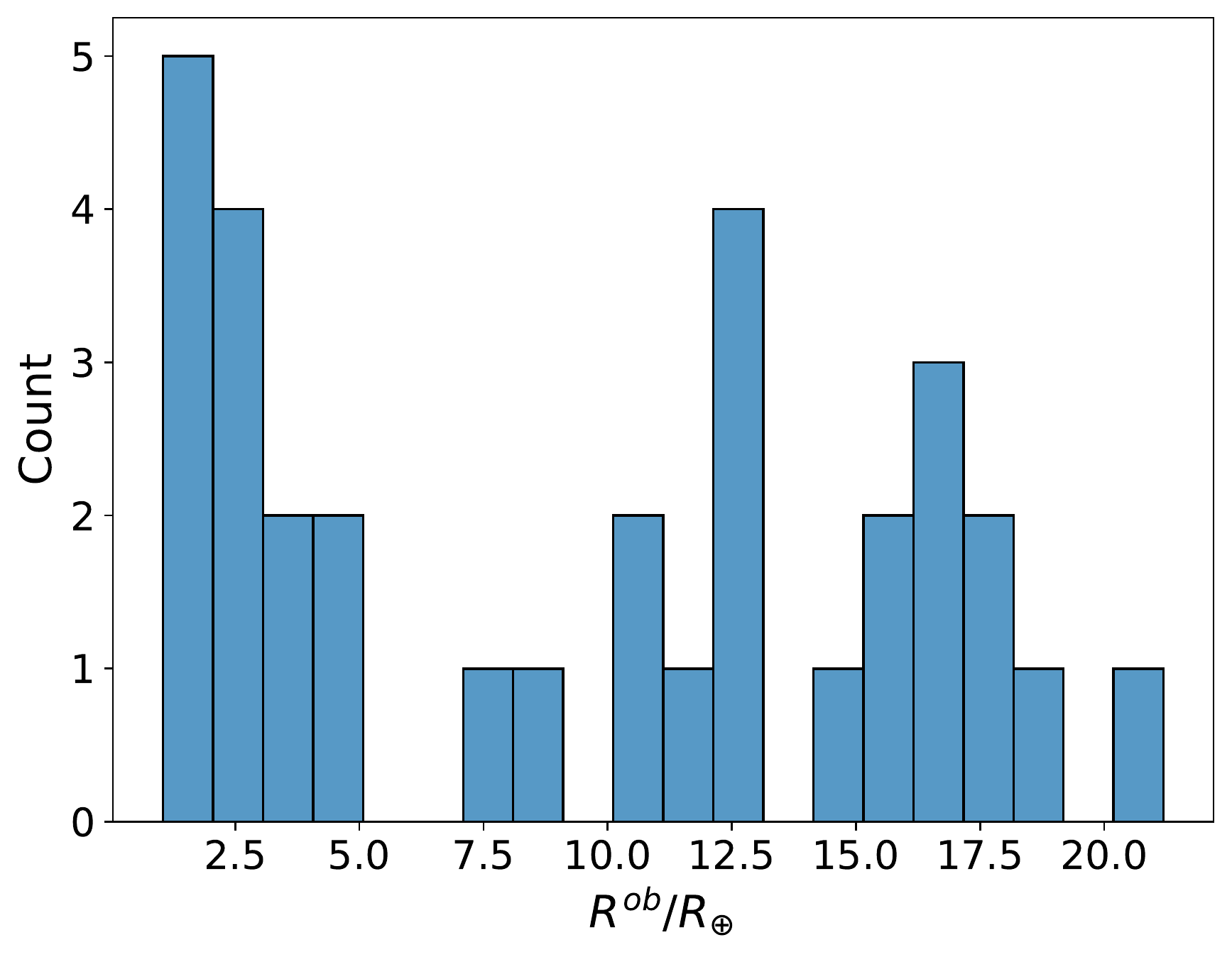}
  \captionof{figure}{Histogram of sample planets whose $\sigma^{Mob}_s$ is larger than 0.207.}
  \label{fig:scaled_Merror}
\end{minipage}%
\begin{minipage}{.5\textwidth}
  \centering
  \includegraphics[width=.85\linewidth]{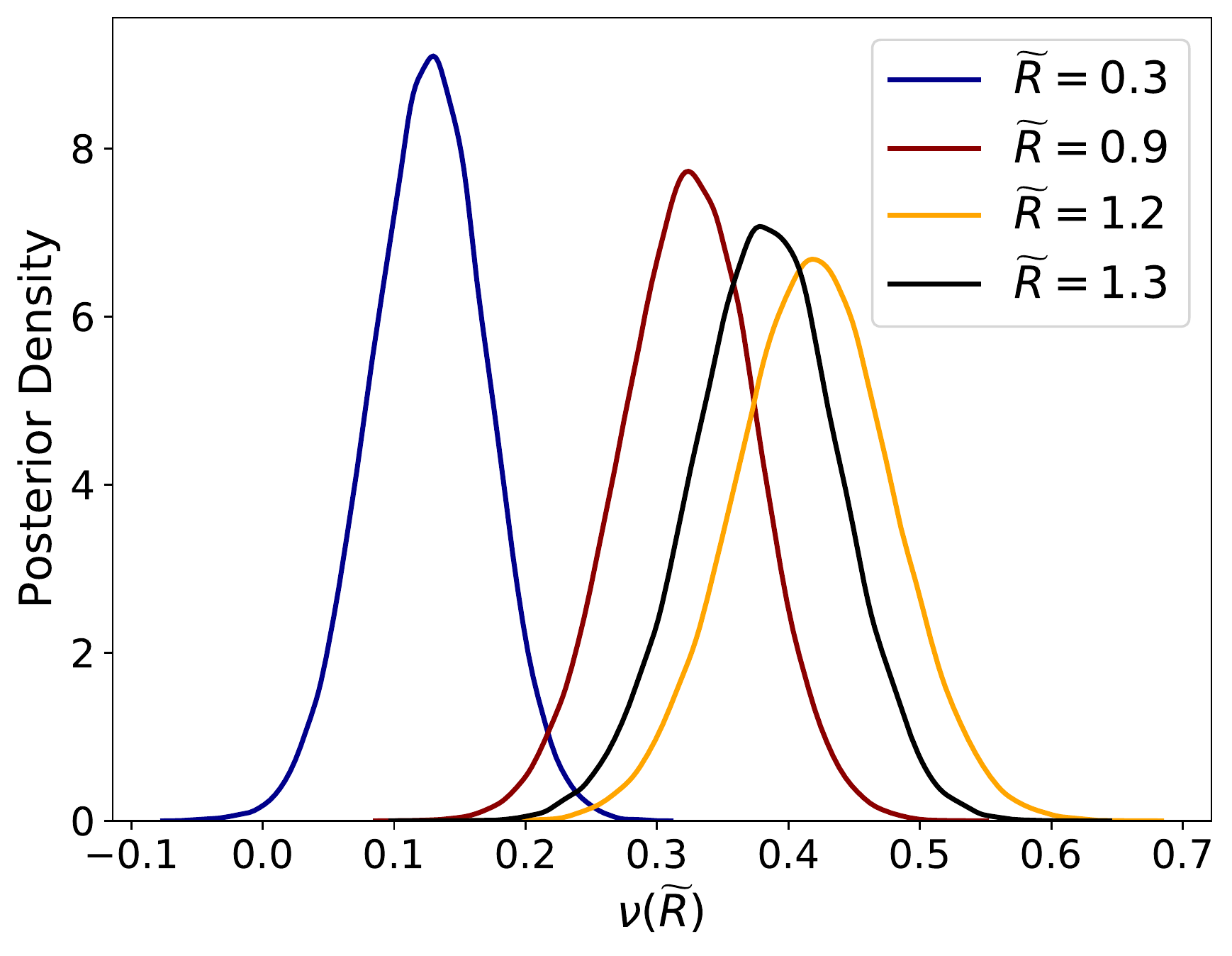}
  \captionof{figure}{Posterior distributions of $\nu(\widetilde{R})$ at different $\widetilde{R}$.}
  \label{fig:nu_pos_atR}
\end{minipage}
\end{figure*}

The Bayesian studentized residuals for this layer, denoted by $\{r^{Mob}_i,i=1,\ldots,n\}$ are plotted in Figure \ref{fig:residual_Mob}. Although the residuals are still distributed around $0$, most of them have absolute values less than 0.5, implying normality assumption leads to overestimation of the observational scatter. A Q-Q plot of these residuals is also provided in Figure \ref{fig:QQ_residual_Mob} to verify such departure from normality. \color{red} \color{black}

% \begin{figure}
% \centering
%   \includegraphics[width=8cm]{QQ_residual_Mob.pdf}
%   \caption{\small
%   Q-Q plot of $r_i^{Mob}$.}
%   \label{fig:QQ_residual_Mob}
% \end{figure}

Therefore, the normality assumption may not be an appropriate choice to model the measurement error. This phenomenon has been hypothesized by other researchers but to the best of our knowledge has not been quantitatively justified as we have illustrated here. We therefore explored the use of the generalized normal (GN) distribution \footnote{$\text{GN}(\mu,\alpha,\beta)$ has the density $p(x;\mu,\alpha,\beta) = \frac{\beta}{2 \alpha \Gamma(1 / \beta)} e^{-(|x-\mu| / \alpha)^{\beta}}$, where $\Gamma(\cdot)$ denotes the gamma function.} \citep{Nad05} to model the residuals $r_i^{Mob}$ and its appears to provide a very well fit as shown in Figure \ref{fig:GN_residual_Mob}, which indicates that the measurement errors can be modeled with a similar form. It thus remains as a part of our future work to find further evidence to support the non-normal assumption and validate it to develop more flexible models for measurement errors.
% \begin{figure}
% \centering
%   \includegraphics[width=8cm]{GN_residual_Mob.pdf}
%   \caption{\small
%   Histogram and empirical fitting curve of $r_i^{Mob}$ using $\text{GN}(-0.06, 0.06, 0.73)$. The fitted parameters are obtained by maximum likelihood estimation.}
%   \label{fig:GN_residual_Mob}
% \end{figure}

We also observe heteroskedasticity in Figure \ref{fig:residual_Mob}, i.e., there are more planets with relatively large studentized residuals in the radius regions smaller than $5R_{\oplus}$ or around $15R_{\oplus}$. Such pattern is in agreement with the distribution of the scaled measurement error of mass $\sigma^{Mob}_s = \sigma^{Mob} / M^{ob}$. For all sample planets, the $90\%$ quantile of $\sigma^{Mob}_s$ is 0.207. In Figure \ref{fig:scaled_Merror}, we plot a histogram of the sample planets whose $\sigma^{Mob}_s$ is larger than this quantile, which shows that the planets are clustered in the aforementioned regions. With a larger measurement error, the estimation of the true mass could be more difficult and the corresponding studentized residual would be therefore larger.
% \begin{figure}
% \centering
%   \includegraphics[width=8cm]{scaled_Merror.pdf}
%   \caption{\small
%   Histogram of sample planets whose $\sigma^{Mob}_s$ is larger than 0.207.}
%   \label{fig:scaled_Merror}
% \end{figure}

\section{Discussion} \label{sec:Discussion}

\subsection{The Impact of Flux on the M-R Relation} \label{subsec:compare}
In this section, we illustrate how the M-R relation depends on the flux. As shown in Equation \ref{eq:mu_M}, the conditional mean of $\widetilde{M}$ is modeled as a linear function of $\widetilde{F}$, and the slope $\nu(\widetilde{R})$ measures how the flux impacts the M-R relation. In Figure \ref{fig:nu_pos_atR}, the posterior distributions of $\nu(\widetilde{R})$ at different $\widetilde{R}$ are plotted. Since all these distributions are away from zero, the impact of the flux is thus nonignorable. Figure \ref{fig:M_pos}(d) summarizes the behavior of $\nu(\widetilde{R})$, indicating that the impact is not uniform along the radius. Particularly, hot-Jupiters with radius around $11R_{\oplus}$ exhibit the strongest dependency on the flux. 

% \begin{figure}
% \centering
%   \includegraphics[width=8cm]{nu_pos_atR.pdf}
%   \caption{\small
%   Posterior distributions of $\nu(\widetilde{R})$ at different $\widetilde{R}$.}
%   \label{fig:nu_pos_atR}
% \end{figure}

\begin{figure*}
\centering
   \includegraphics[width=15cm]{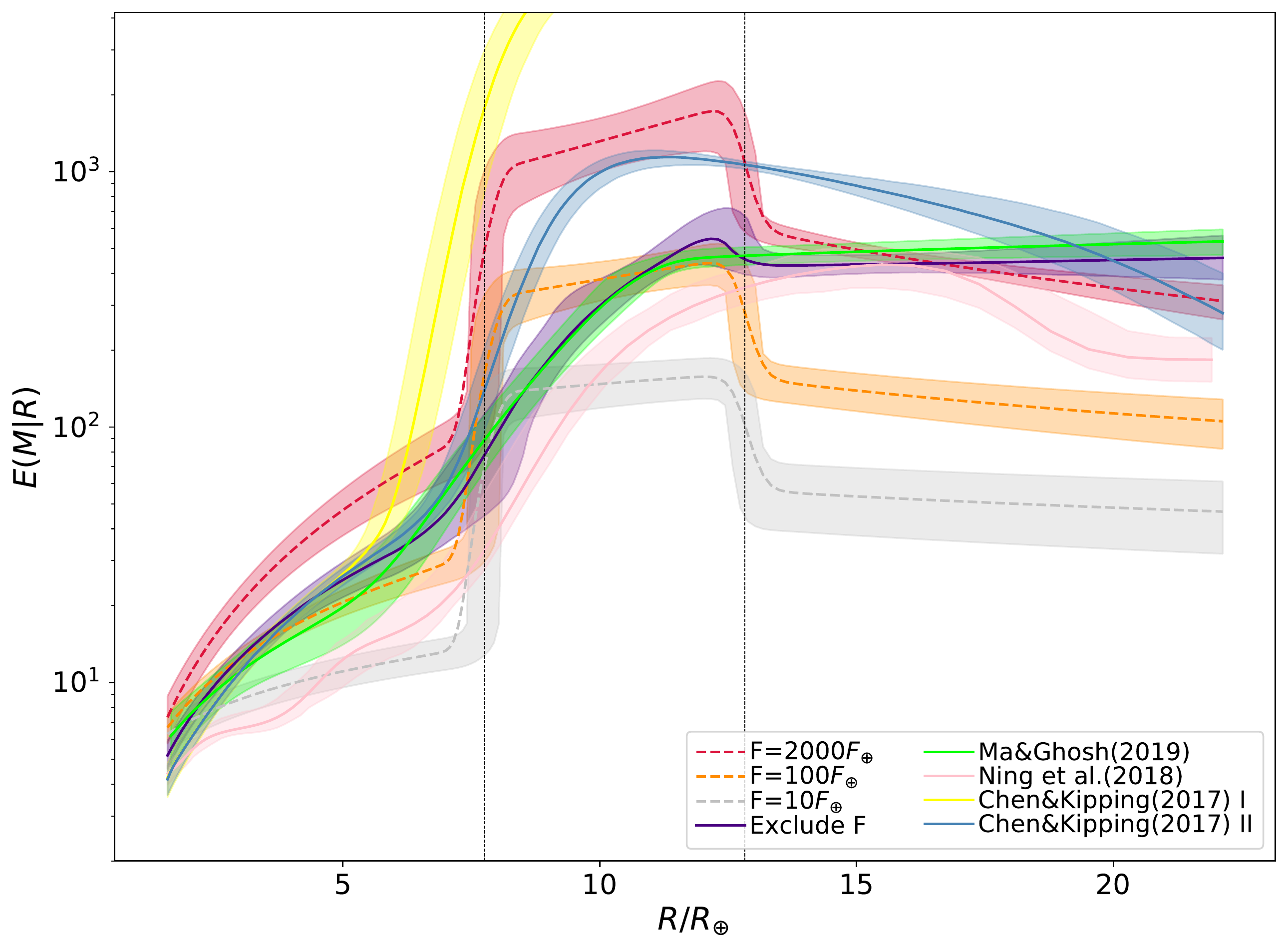}
   \caption{\small
   The mass-radius relation obtained by our model and previous work. The shaded region is the 68\% confidence interval the M-R relation. The blue and yellow curves are obtained from \citet{Chen17} with upper bounds of the log mass grid at $\log_{10}(3\times10^5M_{\oplus})$ and $\log_{10}(13M_J)$, respectively. The yellow curve is horizontally truncated as the range of $y$ axis is intentionally limited to better display the details of other M-R relations.}
   \label{fig:MR_compare}
\end{figure*}

To further illustrate the effect of flux, we plot the M-R relation from our model under different levels of flux using dashed curves in Figure \ref{fig:MR_compare}. We also refit our model on a \textit{modified} data set where the sample planets all have fixed flux. Specifically, $F^{ob}$ is set to be $1000F_{\oplus}$ that is most common in our data set (see Figure \ref{fig:Fob_marginal}), and $\sigma^{Fob}$ is set to be a small value (i.e. $0.1F_{\oplus}$). In this way, we exclude the effect of flux from our model and plot the corresponding M-R relation using the purple curve \footnote{We also provide another way to marginalize the flux in Appendix \ref{app:kde}.}. We also plot the M-R relations obtained by previous works in Figure \ref{fig:MR_compare}. The method of estimating $E(M|R)$ from the results of \citet{Chen17} that attempt to model $E(\widetilde{R}|\widetilde{M})$ is described in Appendix \ref{app:ck}.

As shown in Figure \ref{fig:MR_compare}, the M-R relation obtained by \citet{Ma19} almost overlaps with that from our model after excluding the effect of the flux, which is as expected since they also adopt the broken power law and have a similar sample of planets. The M-R relation from \citet{Nin18} exhibits a similar pattern but consistently prefers lower mass possibly because they use a much smaller sample with fewer massive planets. Compared to our M-R relation that accounts for the impact of the flux, these two M-R relations tend to underestimate the mass for planets with higher flux and smaller radius ($\lessapprox 13R_{\oplus}$). On the other hand, for planets with lower flux, they overestimate the mass along the entire radius range. 

As mentioned in Appendix \ref{app:ck}, the M-R relation obtained by \citet{Chen17} can be largely changed by the upper bound of the log mass grid. Therefore, we plot two M-R relations with different upper bounds. As shown by the yellow curve, with the upper bound at $\log_{10}(3\times10^5M_{\oplus})$ used by \citet{Chen17}, the mean mass increases rapidly along the radius and becomes larger than all the others after $R=6R_{\oplus}$. It is because they use a sample including astronomical objects with mass up to $0.87M_{\odot}$. With the continuity condition applied to the broken power law, the mean mass (especially for sub-Saturns and Jupiters) is pushed upwards by the impact of the included brown dwarfs and low-mass stars. Such impact could be mitigated with a smaller upper bound. For illustration, the blue curve denotes the M-R relation obtained with an upper bound at $\log_{10}(13M_J)$, which agrees with some other M-R relations for smaller planets, but still tends to overestimate for larger planets.

\subsection{Predict Masses Using the M-R-F Relation}
The prediction of mass of a planet is critical for radial velocity surveys. For example, the TESS mission \citep{Ric14} specially designed for small planets transiting small stars has discovered 581 candidates with radius less than $4R_{\oplus}$ as of April 20th, 2020. To schedule the resource-intensive radial velocity campaign towards those potentially habitable exoplanets, the accurate mass prediction of them would be of great importance for assessing their detectability. Compared to the methods of predicting masses solely depending on the M-R relations, the use of insulation flux as an additional object could reduce the intrinsic scatter and therefore likely to yield more accurate predictions.

\begin{figure}
\centering
   \includegraphics[width=8cm]{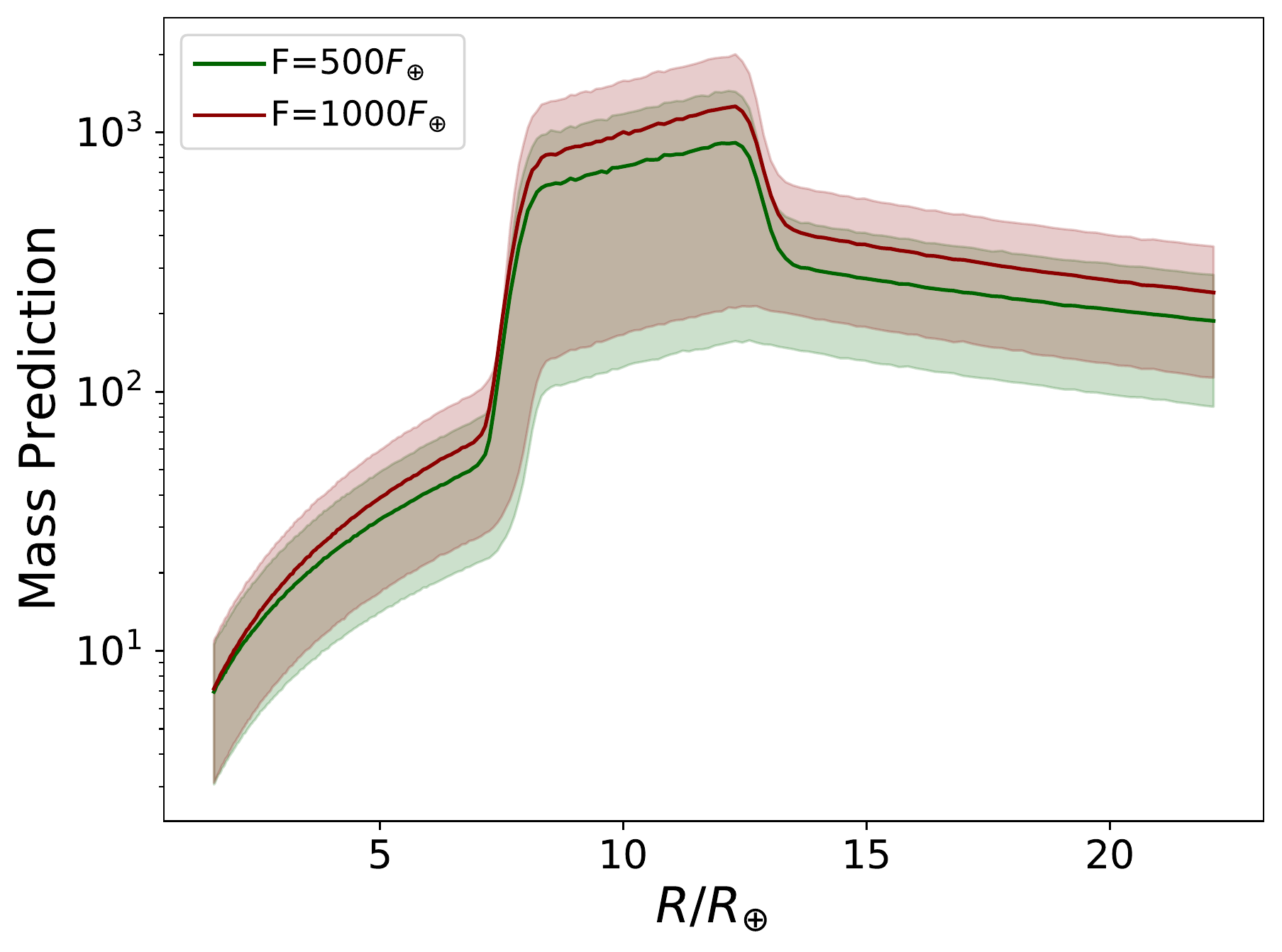}
   \caption{\small
   Posterior mean (solid curves) and 68\% prediction region (shaded areas) of the mass given the radius and the flux.}
   \label{fig:predM_vs_R}
\end{figure}

The prediction of masses based on our modeled  M-R-F relation can be accomplished by the corresponding conditional posterior predictive distribution defined as

\begin{equation}
    p(\widetilde{M}|\widetilde{R}, \widetilde{F}, \mathcal{D}) = \int p(\widetilde{M}|\widetilde{R}, \widetilde{F}, \Phi)p(\Phi|\mathcal{D})d \Phi,
    \label{eq:ppd}
\end{equation}
where $p(\Phi|\mathcal{D})$ denotes the posterior distribution of the parameters in our M-R-F model. Using the posterior samples of $\Phi$, we plot the posterior 68\% prediction region of mass versus the radius under two levels of flux in Figure \ref{fig:predM_vs_R} that shows higher flux could lead to the larger values of predicted mass at the same radius value.

As an example, we calculate the mass prediction of HR858 b/c/d and TOI-813 b that are all exoplanet newly confirmed by TESS. Table \ref{tab:mass_prediction} summarizes the measured radii and fluxes of them.

Discovered by \citet{Van19}, HR858 b, c and d are super-Earths orbiting around a naked-eye F-type star. Following the treatment for data preprocessing used by \cite{Chen17}, their mass predictions from our M-R-F model are computed and also included in Table \ref{tab:mass_prediction}. From limited observations of radial velocities, \citet{Van19} conclude that the 95\% upper limit on the masses of the three planets is around $45M_{\oplus}$ using the RadVel package\citep{Ful18}. For comparison, our estimates of the same quantity are about $30M_{\oplus}$, $26M_{\oplus}$ and $27M_{\oplus}$ for HR858 b, c and d respectively, which are in good agreement with their estimate.

TOI-813 b is a Neptune-like exoplanet discovered by \citet{Eis20}. It is a transiting planet orbiting around an evolved star with a long orbital period, and is therefore of great interest for RV follow-ups. Using the M-R model developed by \cite{Chen17}, \citet{Eis20} report a mass prediction at $42^{+49}_{-19}M_{\oplus}$. Considering that our model also incorporates the flux and is conditioned on a more comprehensive exoplanet sample, our predicted mass at around $14^{+12}_{-6}M_{\oplus}$ has smaller uncertainty and could be more accurate.

%\textcolor{blue}{[The predicted masses of HR858 b/c/d are within the upper limit computed using radial velocity by the author. It may make our prediction of TOI-813 b more convincing.]}

\begin{table}
\centering
\caption{Mass predictions (68\% central prediction interval) of newly confirmed exoplanets by TESS.}
\label{tab:mass_prediction}
\begin{tabular}{cccc}
\hline
Planet Name & $R^{ob} / R_{\oplus}$ & $F^{ob} / F_{\oplus}$ & Mass Pred./$M_{\oplus}$\\
\hline
HR858 b & $2.085^{+0.068}_{-0.064}$ & $989.70_{-55.80}^{+62.50}$ & $8.72_{-4.03}^{+7.48}$  \\
HR858 c & $1.939^{+0.069}_{-0.069}$ & $512_{-26}^{+29}$ & $7.49_{-3.48}^{+6.39}$  \\
HR858 d & $2.164^{+0.086}_{-0.083}$ & $217_{-12}^{+13}$ & $7.98_{-3.65}^{+6.76}$  \\
TOI-813 b & $6.71^{+0.38}_{-0.38}$ & $23.1^{+4.6}_{-3.1}$ & $13.86^{+12.32}_{-6.43}$  \\
\hline
\end{tabular}
\end{table}

\subsection{Transition Points in Radius}
Our M-R-F model identifies two transition locations in radius at around $8R_{\oplus}$ and $13R_{\oplus}$ that divide the radius space into Neptunes, sub-Saturns and Jupiters. These are slightly larger than those found by earlier researchers which explored only the mass-radius space \citep[e.g.][]{Ma19, Nin18, Bas17}. The masses of the planets in the first two regions both increase in radius, as the degeneracy pressure only plays a minor role in determining the radius for less massive planets \citep{Zap69}. However, as shown in Figure \ref{fig:M_pos}(b), the intrinsic scatter of sub-Saturns is significantly higher than that of Neptunes. A possible explanation is that sub-Saturns have larger variation in the fraction of H/He envelope, and the equation of state of these light elements significantly impact the observed M-R-F relation. The second transition point could be treated as the threshold beyond which the compression due to the large mass starts to take strong effect, and significant mass loss due to high flux might occur.

\subsection{Selection Effects}
The selection effects persisting with any piratical sample of exoplanets have two major sources. The first is the non-constant detection completeness that is decided by a combination of factors including the instrument and the data processing pipeline. For example, transit surveys tend to detect planets with larger radius and higher incident flux. This issue can be partially addressed by obtaining a more homogeneous sample \citep[e.g.][]{Wol16, Nei18, Nei20}. We instead don't constrain our sample as the mixture model has better capability in handling heterogeneity. One can also corrects the detection bias by modeling the survey completeness as a function of parameters of interest \citep[e.g.][]{Ful17, Nei20}, which is the direction of our future work.

The ground-based follow-up observations also introduce selection bias that is much harder to deal with. Unlike the detection completeness that becomes less concerning when estimating the conditional distribution (e.g. $f(\widetilde{M}| \widetilde{R}, \widetilde{F})$), the follow-up strategy could bias the estimation of both joint and conditional distribution in the same manner. Meanwhile, the decision process is usually not transparent and subject to human evaluation, which makes it impractical to model the selection function quantitatively for existing catalogs. To fully address this follow-up selection bias in subsequent statistical modeling, it is critical for the follow-up groups to report their selection function in a tractable way as well as all non-detections \citep{Bur18, Mon18}.

\section{Conclusion} \label{sec:Conclusion}
In this work, we present a Bayesian hierarchical finite mixture model (BFMM) to approximate the 3-dimensional joint distribution of the planetary mass, radius and flux. Conditioned on a sample of 319 exoplanets, the key findings from our models are summarized below:

\begin{itemize}
    \item The relationship between mass and radius has a nonnegligible dependence on the flux, especially for hot-Jupiters with radius around $11R_{\oplus}$. The planets receiving higher level of flux tend to be denser, possibly because of the H/He envelope evaporation caused by stellar XUV flux. Hot-Jupiters larger than $13R_{\oplus}$ and receiving strong stellar irradiation exhibit a trend of decreasing mass with increasing radius, indicating that significant atmospheric mass loss could happen during their evolution.
    \item With the assumption of broken power law, we find two transition locations in radius at around $8R_{\oplus}$ and $13R_{\oplus}$, which are slightly larger than those found by previous works \citep[e.g.][]{Ma19, Nin18, Bas17}.
    \item The flux is a key ingredient for mass prediction. The M-R relation that fails to account for the flux may overestimate or underestimate the mass for planets with low or high flux, respectively.
\end{itemize}

It is to be noted that our proposed modeling framework that make use of FMM, can also be adopted to explore the impact of period (P) on M-R relation and can possibly be extended to develop more general models that can approximate the 4-dimensional joint distribution of (Mass)M-Radius(R)-Flux(F)-Period(P). However, such higher-dimensional extensions are admittedly non-trivial as we'd need a much larger sample of data set on M-R-F-P to accurately estimate such a 4-d distribution.

From a methodological perspective, a major contribution of this work is that we proposed two novel methods for model checking which can be used more broadly than just for exploring the validity of our proposed joint models. The robust Hotelling's test can be used to measure the discrepancy between the model and the observed data, as well as identify outliers for further examination. And the Bayesian studentized residual analysis is a powerful tool to validate distributional assumptions in Bayesian hierarchical modeling. Finally, we also point out the possible violation of the normality assumptions for the measurement error models that are predominantly used in astronomy literature. We have suggested the use of generalized normal models which requires further more in-depth explorations in future at a more computational cost.

\section*{Acknowledgements}
This paper includes data collected by the \textit{Kepler} mission. Funding for the \textit{Kepler} mission is provided by the NASA Science Mission directorate. This paper makes use of data from the first public release of the WASP data \citep{But10} as provided by the WASP consortium and services at the NASA Exoplanet Archive, which is operated by the California Institute of Technology, under contract with the National Aeronautics and Space Administration under the Exoplanet Exploration Program.

\section*{DATA AVAILABILITY}
The data and code underlying this article are available in Zenodo, at \url{https://doi.org/10.5281/zenodo.4774442}. The data set was derived from sources in the public domain: The Confirmed Planets table of NASA Exoplanet Archive, at \url{https://dx.doi.org/10.26133/NEA1}.

%%%%%%%%%%%%%%%%%%%% REFERENCES %%%%%%%%%%%%%%%%%%

\bibliographystyle{mnras}
\bibliography{MRF} % if your bibtex file is called example.bib
%%%%%%%%%%%%%%%%%%%%%%%%%%%%%%%%%%%%%%%%%%%%%%%%%%

\appendix

\section{Remove the impact of flux using Kernel density estimator}\label{app:kde}
In Section \ref{subsec:compare}, the impact of flux is removed by refitting our model on a modified data set. Here we show another way to achieve the same purpose based on the kernel density estimator (KDE):

\begin{enumerate}
    \item Using a set of model parameters $\Theta$ and following Equation \ref{eq:mix_F}, \ref{eq:mix_R} and \ref{eq:mix_M}, generate flux, radius and mass samples $\{(F_d, R_d, M_d), d=1,\ldots, D\}$, where $D$ is the number of samples.
    
    \item Use KDE and $\{(R_d, M_d), d=1,\ldots, D\}$ to estimate the joint distribution of mass and radius, denoted by $\widehat{p}(R,M;\Theta)$.
    
    \item For a given radius $R$, Estimate $E(M|R)$ with

$$
\widehat{E}(M|R;\Theta) = \sum_d \widehat{p}(M_d|R)M_d = \sum_d \frac{\widehat{p}(M_d,R)}{\sum_d\widehat{p}(M_d,R)}M_d
$$
    \item Repeat (i) to (iii) for $S$ sets of posterior model parameters to get $\{\widehat{E}(M|R;\Theta^{(s)}), s=1, \ldots, S\}$ whose average and quantiles are the posterior estimate and credible interval of $E(M|R)$.
\end{enumerate}

\begin{figure}
\centering
   \includegraphics[width=7cm]{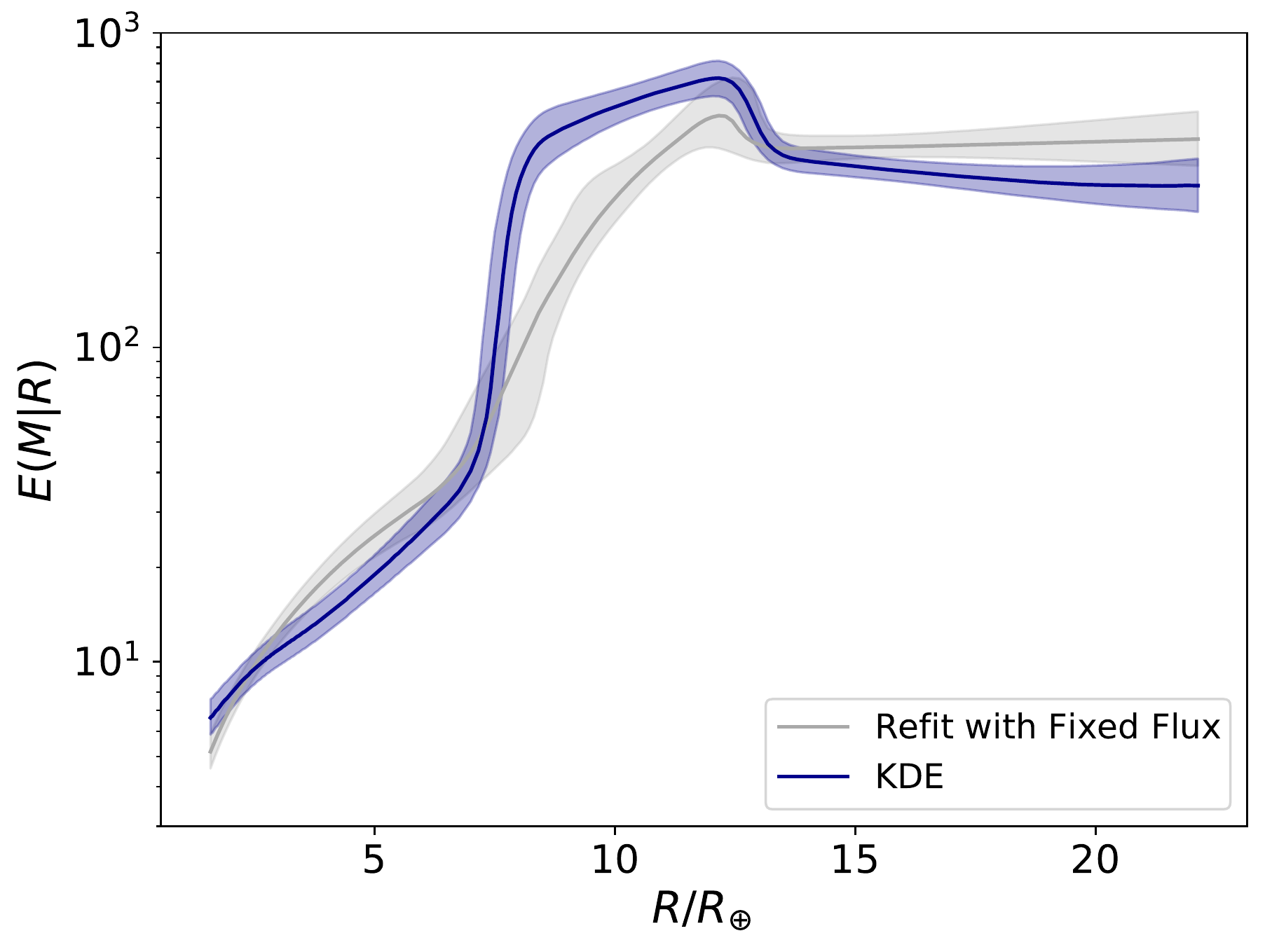}
   \caption{\small
   Comparison of two methods to exclude the impact of flux from our model.}
   \label{fig:exclude_compare}
\end{figure}

In Figure \ref{fig:exclude_compare}, we compare the M-R relations obtained by the two methods of removing the impact of flux. In general, they follow a similar trend but exhibit slight difference around $10R_{\oplus}$. It's because the model parameters used for sample generation are still subject to the impact of flux, although we attempt to marginalize the flux later. In our words, the method of refitting the model with fixed flux removes the impact of flux more completely.

\section{From Conditional Expectation of Radius to Conditional Expectation of Mass }\label{app:ck}
Different from other works included in Figure \ref{fig:MR_compare}, \citet{Chen17} model $E(\widetilde{R}|\widetilde{M})$, and thus we cannot obtain $E(M|R)$ directly from their results. In this section, we describe how to deal with it following their treatment for mass prediction (see Section 5.3 in \citet{Chen17}).

For a given radius $R$ (or $\widetilde{R}$), the expected mass can be estimated with the following steps:
\begin{enumerate}
    \item Prepare an equally partitioned grid of mass in log scale, denoted by $\{\widetilde{M}_{\text {g }}^{(q)}, q=1,\ldots, Q\}$, where $Q$ is the number of grid points, and $\widetilde{M}_{\text {g }}^{(Q)}$ is the upper bound of this grid.
    
    \item Using a set of model parameters $\Theta$, Estimate
$$
P(\widetilde{M}_{\text {g }}^{(q)}|\widetilde{R};\Theta) \propto P(\widetilde{R}|\widetilde{M}_{\text {g }}^{(q)};\Theta)P(\widetilde{M}_{\text {g }}^{(q)}) \;\; q=1\ldots Q
$$
    where $P(\widetilde{R}|\widetilde{M};\Theta)$ is directly modeled by \citet{Chen17}, and $P(\widetilde{M})$ is a constant as $\widetilde{M}$ has a uniform prior in their model.
    
    \item Perform weighted sampling from the grid of mass with the above probabilities. Let $\{\widetilde{M}_d, d=1,\ldots,D\}$ denote the sampled log masses. Then $E(M|R)$ can be estimated by 
$$
\widehat{E}(M|R;\Theta) = \frac{1}{D}\sum_{d=1}^D 10^{\widetilde{M}_d}
$$
    \item Repeat (i) to (iii) for $S$ sets of posterior model parameters to get $\{\widehat{E}(M|R;\Theta^{(s)}), s=1, \ldots, S\}$ whose average and quantiles are the posterior estimate and credible interval of $E(M|R)$.
\end{enumerate}

We find that the choice of the upper bound $\widetilde{M}_{\text {g }}^{(Q)}$ can largely change the shape of the obtained M-R relation. Therefore, two M-R relations from \citet{Chen17} are plotted, one with their original upper bound at $\log_{10}(3\times10^5M_{\oplus})$, the other with an upper bound at $\log_{10}(13M_J)$.

% Don't change these lines
\bsp	% typesetting comment
\label{lastpage}
\end{document}